\documentclass[aip,notitlepage, nofootinbib, floatfix, reprint, onecolumn]{revtex4-1}%
\usepackage{amssymb}
\usepackage{amsfonts}
\usepackage{amsmath}
\usepackage{graphicx}
\usepackage{enumerate}

\usepackage[usenames,dvipsnames]{color}
\usepackage{ulem}%
 \newcommand{\balpha}{\boldsymbol{\alpha}}
 \newcommand{\bUpsilon}{\boldsymbol{\Upsilon}}
 \newcommand{\bPi}{\overline{\overline{\boldsymbol{\Pi}}}}
 
 \newcommand{\bx}{\mathbf{x}}
 \newcommand{\bX}{\mathbf{X}}
 \newcommand{\br}{\mathbf{r}}
 \newcommand{\bp}{\mathbf{p}}
 \newcommand{\bP}{\mathbf{P}}
 \newcommand{\bR}{\mathbf{R}}
 \newcommand{\bT}{\mathbf{T}}
 \newcommand{\bF}{\mathbf{F}}
 \newcommand{\bL}{\mathbf{L}}
 \newcommand{\bGamma}{\boldsymbol{\Gamma}}
 \newcommand{\bxi}{\boldsymbol{\xi}}
 \newcommand{\bfF}{\boldsymbol{\mathfrak{F}}}
 \newcommand{\bOmega}{\boldsymbol{\Omega}}
 \newcommand{\bomega}{\boldsymbol{\omega}}
 \newcommand{\bPhi}{\boldsymbol{\Phi}}
 \newcommand{\bnabla}{\boldsymbol{\nabla}}
 \newcommand{\bfv}{\boldsymbol{\mathfrak{v}}}
 \newcommand{\bff}{\boldsymbol{\mathfrak{f}}}
 \newcommand{\bfr}{\boldsymbol{\mathfrak{r}}}
  \newcommand{\bfj}{\boldsymbol{\mathfrak{j}}}
 \newcommand{\bfp}{\boldsymbol{\mathfrak{p}}}
 \newcommand{\bfm}{\boldsymbol{\mathfrak{m}}}
 \newcommand{\wbfp}{\widetilde{\boldsymbol{\mathfrak{p}}}}
 \newcommand{\blambda}{\boldsymbol{\lambda}}
 \newcommand{\bLambda}{\boldsymbol{\Lambda}}
 \newcommand{\bgamma}{\boldsymbol{\gamma}}
  \newcommand{\bepsilon}{\boldsymbol{\epsilon}}
 \newcommand{\FnN}{F^{(n+N)}}
 \newcommand{\fN}{f^{(N)}}
 \newcommand{\fn}{f^{(n)}}
 \newcommand{\rhon}{\rho^{(n)}}
 \newcommand{\red}[1]{\textcolor{black}{#1}}

\begin{document}

\title{Dynamical density functional theory for
       orientable colloids including inertia and hydrodynamic interactions}

\author{Miguel A. \surname{Dur\'{a}n-Olivencia}}
\email{m.duran-olivencia@imperial.ac.uk}
\affiliation{\label{ICL}Department of Chemical Engineering, Imperial College London, 
	     South Kensington Campus, London SW7 2AZ, UK}

\author{Benjamin D. \surname{Goddard}}
\email{b.goddard@ed.ac.uk}
\affiliation{\label{SOM} School of Mathematics and the Maxwell Institute for 
	      Mathematical Sciences, University of Edinburgh, Edinburgh EH9 3FD, UK}
\author{Serafim Kalliadasis} 
\email[Corresponding author:]{ s.kalliadasis@imperial.ac.uk}
\homepage{http://www.imperial.ac.uk/complex-multiscale-systems/}
\affiliation{\label{ICL}Department of Chemical Engineering, Imperial College London, 
	     South Kensington Campus, London SW7 2AZ, UK}
\date{\today}

\begin{abstract}
Over the last few decades, classical density-functional theory (DFT) and
its dynamic extensions (DDFTs) have become powerful tools in the study of
colloidal fluids. Recently, previous DDFTs for spherically-symmetric
particles have been generalised to take into account both inertia and
hydrodynamic interactions, two effects which strongly influence
non-equilibrium properties. The present work further generalises this
framework to systems of anisotropic particles. Starting from the Liouville
equation and utilising Zwanzig's projection-operator techniques, we derive
the kinetic equation for the Brownian particle distribution function, and
by averaging over all but one particle, a DDFT equation is obtained. Whilst
this equation has some similarities with DDFTs for
spherically-symmetric colloids, it involves a translational-rotational
coupling which affects the diffusivity of the (asymmetric) particles. We
further show that, in the overdamped (high friction) limit, the DDFT is
considerably simplified and is in agreement with a previous DDFT for
colloids with arbitrary shape particles. 
\end{abstract}

\maketitle

\section{Introduction}\label{intro}

The study of colloidal fluids, which typically involve particles of
micrometer size suspended in a simple fluid bath (where the particles are the
atoms-molecules themselves, hence of nanometer size), goes back to the 19th
Century with the work of \citet{brown_brief_1828}. Since then, numerous
studies have been devoted to the description of the dynamics of such systems.
A particular challenge is the multiscale nature of the dynamics due to the
mass separation of the bath and colloidal particles which in turn implies
that the velocities of the bath particles are much higher than those of the
colloidal particles. While the time evolution of colloidal fluids can be
formally studied by modelling the full systems, i.e. considering the
Newtonian dynamics of both bath and colloidal particles, such an approach is
computationally intractable, due to both the number of particles and the very
different timescales one has to
consider\cite{goddard_general_2012,goddard_unification_2013}, and thus there
is a need for coarse graining leading to reduced models.

One such reduced model consists of the stochastic Langevin Equations (LEs)
for colloidal positions and momenta or equivalently the corresponding
Fokker-Planck equation (FPE) for the $N$-body probability distribution
function. The LEs were originally proposed heuristically, but can be formally
justified as shown in Appendix \ref{app:projection-operator-tech} (where for
the first time a detailed microscopic derivation of the rotational part of
the FPE is offered starting from the full system of bath and colloidal
particles).
A consequence of this coarse graining down to the dynamics of only the
colloidal-particle degrees of freedom is that we no longer deal with
deterministic but with random variables. The advantage is that it removes the
enormous number of variables related to the dynamical variables of the bath,
traditionally via projection-operator techniques which effectively allow to
average out the bath\cite{kirkwood_statistical_1946, murphy_brownian_1972,
deutch_molecular_1971, wilemski_derivation_1976, ermak_brownian_1978,
grabert_microdynamics_1980,darve_computing_2009}. Despite such a
simplification, the resultant equations are still computationally prohibitive
because of the large number of particles that need to be taken into
consideration to describe the behavior of colloidal fluids.

A theoretical way out is to obtain mean field approximations which are
independent of the number of particles. For an $N$-particle system, a
standard procedure in the statistical mechanics of classical fluids is to
average over $N-n$ (with $n < N$) colloidal particles, resulting in a
time-evolution equation for the $n$-body reduced distribution function.
However, to obtain a closed set of equations for a given $n$ requires
knowledge of the relationships between the full and the reduced
distributions.
Dynamical density functional theories (DDFTs) perform this procedure for
$n=1$, i.e.\ for the one-body density, and typically result in a continuity
equation for the density and a time-evolution equation for the current, that
is now a functional of the density. 
%
The functional that relates the current and the density is generally unknown,
but, for a practical implementation of DDFTs it needs to be approximated. A
standard approach, for the overdamped dynamics, is to take it as the
free-energy functional of a system with the same density at equilibrium. This
functional has been well-studied in the statistical mechanics of
fluids\cite{evans_nature_1979,lutsko_recent_2010}. Also, approximating the
functional relating the current and the density with the free energy at
equilibrium also ensures that the DDFTs reduce to the corresponding
equilibrium DFTs\cite{goddard_general_2012,goddard_unification_2013}.

The first derivations of a DDFT go back to the early works of
\citet{evans_nature_1979} and \citet{dieterich_nonlinear_1990}. These initial
formulations were mostly phenomenological and subsequent efforts focused on
deriving rigorously the same results and also DDFTs that include additional
effects such as inertia and hydrodynamic interactions
(HIs)\cite{marconi_dynamic_2000,archer_dynamical_2004,archer_dynamical_2009,espanol_derivation_2009,goddard_overdamped_2012,goddard_general_2012,goddard_unification_2013,donev_dynamic_2014}.

There have also been several efforts to test the derived DDFTs and their
predictive capabilities. In fact there have been successful applications to a wide
spectrum of physical settings, from hard rod and hard sphere
systems\cite{goddard_unification_2013},
mixtures\cite{goddard_multi-species_2013}, thermodynamic phase transitions such
as nucleation and spinodal
decomposition\cite{van_teeffelen_colloidal_2008,lutsko_dynamical_2012} to the
calculation of the van Hove distribution function for Brownian hard
spheres\cite{hopkins_van_2010} and
crystallization\cite{neuhaus_density_2014}.

DDFTs have been derived for both the overdamped and inertial regimes and
although they were initially applied to spherically-symmetric colloids, they
have also been generalised to systems where both orientational and
translational motions of particles are taken into account, either under weak\cite{rex_dynamical_2007} or strong\cite{wittkowski_dynamical_2011}
coupling between the two motions. However, these generalisations are
restricted to the overdamped regime and neglect HIs. The main goal of the
present study is precisely to formulate a DDFT for orientable particles to
include both inertia and HIs. Our methodology follows closely the
momentum-moments approach adopted by~\citet{archer_dynamical_2009}
to obtain a DDFT for spherically-symmetric colloidal particles with inertia and also
by~\citeauthor{goddard_general_2012}\cite{goddard_general_2012,goddard_unification_2013} to include both inertia and HIs in this DDFT.

In Section \ref{sec:evolution-equations} we introduce the equations of motion
for a system of $N$ interacting orientable colloidal particles immersed in a
thermal bath. Analogous to the spherically-symmetric case, these equations
are the LEs for the phase-space coordinates: positions, Eulerian angles,
linear and angular (also referred to as rotational) momenta. These equations
have already been derived and examined thoroughly
before\cite{dickinson_brownian_1985, wolynes_dynamical_1977,
hernandez-contreras_brownian_1996}, but for the sake of clarity and
completeness they are reviewed in Appendices
\ref{app:projection-operator-tech} and \ref{app:neglecting-bath-inertia}, in
particular highlighting their range of applicability. The derived LEs
naturally contain the full rotational-translational coupling which is crucial
in the DDFT derivation that follows.
In Section \ref{sec:continuity-current-equation}, we
follow\cite{goddard_general_2012,goddard_unification_2013} and integrate the Kramer's
equation over all but one particle, yielding a continuity equation for the
density distribution and a time-evolution equation for the current. At this
point we require closure approximations to obtain a final equation dependent
on the density distribution only.
These approximations, discussed in detail in
Section~\ref{sec:extended-unified-DDFT}, are: (i) the local-equilibrium
approximation for the one-body distribution function; (ii) the adiabatic
approximation involving the equilibrium free-energy functional; (iii)
separation of the $N$-body distribution into local equilibrium and
non-equilibrium contributions; and (iv) an \red{Enskog-like} approximation for the
two-body interactions.
We then arrive at the central result of this work, i.e. a generalised DDFT
equation to describe systems of orientable particles taking account of both
inertia and HIs.
Finally, we compare our derived DDFT with previous DDFTs. As expected, if the
rotational degrees of freedom are ignored, our DDFT reduces to the one
obtained for point-like or spherical particles in the presence of both
inertia and HIs\cite{goddard_general_2012,goddard_unification_2013}, and
hence also to several other DDFTs\cite{marconi_dynamic_2000,
archer_dynamical_2009, archer_dynamical_2004, espanol_derivation_2009,
goddard_overdamped_2012} which have neglected inertia or HIs or both.
For orientable particles, neglecting inertial effects and HIs allows us to
connect our DDFT with previous ones, recovering in particular the DDFTs
derived by \citet{rex_dynamical_2007} and \citet{wittkowski_dynamical_2011}.
Concluding remarks along with a discussion of open problems are offered in
Section \ref{sec:conclusions}.

\section{Evolution equations}
\label{sec:evolution-equations}

Consider a closed system of $N$ identical, asymmetric colloidal particles
with mass $m$ immersed in a fluid of $n\gg N$ identical bath particles with
mass $m_b \ll m$. This mass separation plays a vital role in deriving the
Brownian equations of motion governing colloidal dynamics via
projection-operator techniques\cite{kirkwood_statistical_1946,
murphy_brownian_1972, deutch_molecular_1971, wilemski_derivation_1976,
ermak_brownian_1978, grabert_microdynamics_1980,darve_computing_2009} (see
also Appendix \ref{app:projection-operator-tech}).
We start by describing the configuration state of the colloidal particles by
the position vectors $\mathbf{r}^N=\{\mathbf{r}_1,\dots,\mathbf{r}_N\}$ for
their centres of mass and the Eulerian angles
$\balpha^N=\{\balpha_1,\dots,\balpha_N\}$, with
$\balpha_i=(\theta_i,\phi_i,\chi_i)$ defined in the same manner as
\citet{condiff_brownian_1966, goldstein_classical_2002} or
\citet{jose_classical_2013}. These angles characterise the orientation of
principal-axes frame
of the particles,
$\mathfrak{B}$, relative to the space-fixed frame, $\mathfrak{S}$.
We also denote the translational and angular momenta as
$\bp^N=\{\bp_1,\dots,\bp_N\}$ and $\bL^N=\{\bL_1,\dots,\bL_N\}$,
respectively. The angular velocity of the $i$th colloidal particle,
$\boldsymbol{\omega}_i$, is related to its angular momentum by the equation
$\bL_i=\mathbb{I}_i\boldsymbol{\omega}_i$ with $\mathbb{I}_i$ the inertia tensor.
In the principal-axes frames $\mathfrak{B}$, such a tensor is diagonal and
constant, $\mathbb{I}_i\equiv\mathbb{I}=\text{diag}(I_{1},I_{2},I_{3})$.
The transformation between the frames $\mathfrak{B}$ and $\mathfrak{S}$ is given by the rotation operator $\mathcal{R}_i$,\cite{jose_classical_2013} such that $\mathbf{x}=\mathcal{R}_i\,\mathbf{x}'$ where
$\mathbf{x}\in\mathfrak{B}$ and $\mathbf{x}'\in\mathfrak{S}$.
Hence $\boldsymbol{\omega}_i=\mathcal{R}_i\boldsymbol{\omega}_i'$, with $\boldsymbol{\omega}_i'$ the angular velocity in $\mathfrak{S}$, and $\mathbb{I}_i'=\mathcal{R}_i^\top\,\mathbb{I}\,\mathcal{R}_i$, which is neither diagonal nor constant.
The angular velocity $\bomega_i$ can be related to the Eulerian velocities,
$\dot{\balpha}_i$, via the transformation $\bomega_i(\balpha_i) \equiv
\dot{\bPhi_i}(\balpha_i) = \bLambda_i^\top \dot{\balpha}_i$, where the dot
denotes time derivative and $\bPhi_i$ highlights the fact that $\bomega_i$ is
the vector accounting for the rate of change of angular displacement over the
Cartesian frame $\mathfrak{B}$.
The exact
definition of $\bLambda^\top$ can be found in Appendix
\ref{app:projection-operator-tech}.

For simplicity, we will consider angular quantities with respect to the
body-fixed frames. As such, the dynamics of colloidal particles are described
by the Langevin equations\cite{dickinson_brownian_1985,
dickinson_brownian_1985-1} (see detailed derivation in Appendix
\ref{app:projection-operator-tech})
\begin{align}
 \dot{\br}_j(t)=&\,\frac{1}{m}\bp_j,\quad
 \dot{\bPhi}_j(t)=\,\mathbb{I}^{-1}\bL_j,\label{eq:1}\\
 \dot{\bp}_j(t)=&\, -\frac{\partial}{\partial \br_j}V(\br^N,\balpha^N)-\sum_{k=1}^N\left(\bGamma_{jk}^{TT}\bp_k+\bGamma_{jk}^{TR}\bL_k\right)+\sum_{k=1}^N \left( \mathbf{A}_{jk}^{TT}\mathbf{f}_k(t)+\mathbf{A}_{jk}^{TR}\mathbf{t}_k(t) \right),
 \notag
 \\
 \dot{\bL}_j(t)=&
 -\frac{\partial}{\partial \bPhi_j}{V}(\br^N,\balpha^N)-(\mathbb{I}^{-1}\bL_j)\times\bL_j
 \,
 -\sum_{k=1}^N\left(\bGamma_{jk}^{RT}\bp_k+\bGamma_{jk}^{RR}\bL_k\right)
 +\sum_{k=1}^N \left( \mathbf{A}_{jk}^{RT}\mathbf{f}_k(t)+\mathbf{A}_{jk}^{RR}\mathbf{t}_k(t) \right) \notag,
\end{align}
or, more compactly,
\begin{align}
 \dot{\bfr}_j(t)=&\ \bfm_j^{-1}\bfp_j(t)\label{eq:2},\\
 \dot{\bfp}_j(t)=&\ \bfF_j-\sum_{k=1}^N\bGamma_{jk}(\bfr^N)\bfp_k + \sum_{k=1}^N\mathbf{A}_{jk}\boldsymbol{\xi}_k(t)\notag,
\end{align}
with $\bfm_j = \text{diag}(m\mathbf{1},\mathbb{I})$,   $\boldsymbol{\mathfrak{r}}_j=(\mathbf{r}_j,\boldsymbol{\Phi}_j)$,  $\boldsymbol{\mathfrak{p}}_j=(\mathbf{p}_j,\mathbf{L}_j)$ and $\bfF_j=(-\frac{\partial V}{\partial \br_j},-\frac{\partial V}{\partial \bPhi_j}-(\mathbb{I}^{-1}\bL_j)\times\bL_j)$, where $V(\bfr^N)$ is the solvent-averaged interaction-potential of mean force\cite{snook_langevin_2006}.
The gradient operators are defined by
$\boldsymbol{\nabla}_{\mathfrak{r}_j}=(\frac{\partial}{\partial
\mathbf{r}_j},\frac{\partial }{\partial \boldsymbol{\Phi}_j})^\top$, with
$\frac{\partial}{\partial \bPhi}=\mathbf{e}_x\frac{\partial}{\partial
\Phi_x}+\mathbf{e}_y\frac{\partial}{\partial
\Phi_y}+\mathbf{e}_z\frac{\partial}{\partial \Phi_z}$ the angular gradient
[see Equations (\ref{eq:a-17})--(\ref{eq:a-18})], and
$\boldsymbol{\nabla}_{\boldsymbol{\mathfrak{p}}_j}= (\frac{\partial
}{\partial \mathbf{p}_j},\frac{\partial }{\partial \mathbf{L}_j})^\top$.
Here $\bxi_j=(\mathbf{f}_j,\mathbf{t}_j)^\top$ is a 6-dimensional Gaussian
white noise vector
where $\mathbf{f}_j$ and $\mathbf{t}_j$ are random forces and torques acting
upon the $j$-th particle, respectively, such that $\langle
\xi_j^a(t)\rangle=0$ and $\langle \xi_j^a(t)\xi_k^b(t')\rangle =
2\delta_{jk}\delta^{ab}\delta(t-t')$. As for the spherical case, the motion
of the colloidal particles induces flows in the bath which in turn result in
forces on the colloids, what we have already referred to in Section~\ref{intro} as
HIs. These HIs are represented by the $6\times6$ friction tensors, $
 \bGamma_{jk}=(\bGamma_{jk}^{\mu\nu})
$ where the superscripts $\mu,\nu\in\{T,R\}$ denote the coupling between the
translational and angular momenta.
%
Furthermore, the interaction between bath and colloidal particles has a random component
representing unpredictable collisions of the bath particles with the
colloidal particles. The strength of these random forces is given by the
$6\times 6$ tensors $\mathbf{A}_{jk}=(\mathbf{A}_{jk}^{\mu\nu})$ which are
coupled to the friction tensors via the fluctuation-dissipation
relation,\cite{wolynes_dynamical_1977,dickinson_brownian_1985}
\begin{align}
 \bfm\,\bGamma_{jk}(\bfr^N) = \beta\sum_{l=1}^N\mathbf{A}_{jl}(\bfr^N)\mathbf{A}_{kl}(\bfr^N)\label{eq:3}.
\end{align}
Here $\beta=1/k_BT$ with $k_B$ the Boltzmann constant and $T$ the
temperature, is assumed to be constant.
%
We define the $6N\times6N$ total-mass tensor $\bfm_{_N}=\text{diag}(\bfm_j)$, along with the $6N\times 6N$ tensors $\bGamma =(\bGamma_{jk})$,
and $\mathbf{A}=(\mathbf{A}_{jk})$, transforming Equation (\ref{eq:3}) to $\bfm_{_N}\bGamma = \beta\mathbf{A}\mathbf{A}^\top$.
Due to the asymmetry of the particles,
the friction and noise matrices are not generally
symmetric. However, as discussed by \citet{brenner_stokes_1964-2} and
\citet{condiff_brownian_1966}, the asymmetric part of $\bGamma_{ij}$ is
exclusively due to nondissipative gyrostatic forces and torques acting on the
particles. If such effects can be neglected\cite{condiff_brownian_1966}, then
one can assume $\bGamma_{jk}^{TR}=(\bGamma_{jk}^{RT})^\top$. The
fluctuation-dissipation relation is then $\mathbf{A} = \sqrt{k_BT \bfm_{_N}
\bGamma}$, analogous to that for spherical particles.

Starting from the Smoluchowski equation, a fluctuating DDFT equation was
derived recently\cite{donev_dynamic_2014}. The authors of this study argued
that any equation of motion that accounts for inertial effects of the
colloidal particles must include the inertia of the fluid bath. Moreover,
they claim that there are only two consistent ways of dealing with this
problem: either \emph{fluctuating hydrodynamics}\cite{hauge_fluctuating_1973}
or the overdamped LEs based on the work of \citet{hinch_application_1975} and
\citet{roux_brownian_1992}. They justify this claim using the estimate
obtained by \citet{roux_brownian_1992}, who also refers to
\citet{hinch_application_1975}, of the typical relaxation time of the bath,
eventually
invalidating the inertial LEs.
Nevertheless, while the inertial equations (\ref{eq:1}) are neither formally
exact nor completely general, it can be shown that there exist very sensible
regimes in which neglecting inertia in the bath is appropriate,
namely when the colloidal-particle density,
$\rho_B$, is much larger than that of the bath particles, $\rho_b$. This is
done in Appendix \ref{app:neglecting-bath-inertia} using dimensional analysis
showing that there is indeed a mass-induced time-scale separation when $\rho_b/\rho_B\ll1$.
Under such circumstances the LEs and the associated
FPE must be applicable, as explicitly pointed out also by
\citet{roux_brownian_1992}.

From the system of stochastic differential equations (SDEs) in Eq.
(\ref{eq:2}), one can obtain\cite{risken_fokker-planck_1996} (see Appendix
\ref{app:projection-operator-tech}) the time-evolution equation of the
phase-space probability density function (PDF), $\fN(\bfr^N,\bfp^N;t)$, to
find each particle $j$ with positions $\br_j,\ \balpha_j$ and momenta
$\bp_j,\ \bL_j$ at time $t$,
\begin{align}
\partial_t \fN(\bfr^N,\bfp^N;t) +&
\sum_{j=1}^N\bnabla_{\bfr_j}\cdot \big(\bfm^{-1}\bfp_j\fN(\bfr^N,\bfp^N;t) \big)+\sum_{j=1}^N\bfF_j\cdot\bnabla_{\bfp_j}\fN(\bfr^N,\bfp^N;t) \label{eq:4}\\
&= \sum_{j,k=1}^N \,\bnabla_{\bfp_j}\cdot\bGamma_{jk}(\bfr^N)\left(\bfp_k+k_BT\bfm\,\bnabla_{\bfp_k}\right)\fN(\bfr^N,\bfp^N;t)\notag.
\end{align}
It is well-known that Equations (\ref{eq:2}) or (\ref{eq:4}) are computationally
tractable only for small systems. This problem can be circumvented to an
extent by assuming a very simple structure for the friction tensor, namely
$\bGamma = \text{diag}(\gamma_{TT}\mathbf{1},\gamma_{RR}\mathbf{1})$. While
such an approximation certainly reduces the numerical complexity [since
determining $\mathbf{A}$ is now $O(N)$ rather than $O(N^3)$], it physically
corresponds to ignoring HIs, thus neglecting crucial physical information.
Another approach to tackle the problem, which is the one we adopt, is to
derive a coarse-grained/mean-field model by averaging over the degrees of
freedom of all but a few particles\cite{wu_density-functional_2007,
chan_time-dependent_2005,
goddard_unification_2013,goddard_multi-species_2013}.  This method yields a
lower-dimensional system but requires knowledge of the functional relation
between $\fN$ and the reduced distribution function. Therefore, convenient
and accurate
closures for the dynamics of the reduced distribution function must be found.

\section{Continuity and flux equations \label{sec:continuity-current-equation}}

Given (\ref{eq:4}),  we can reduce to a lower dimensional problem by
averaging over all but a few particles.
Our first assumption involves the solvent-averaged
interaction potential $V(\bfr^N)$. Inspired by the decomposition proposed for
spherical particles\cite{goddard_unification_2013} we assume that this
potential can be split into linear combinations of $n$-particle interactions,
i.e.
\begin{align}
 V(\bfr^N;t)= \sum_{j=1}^NV_1(\bfr_j;t)+\sum_{n=2}^N\frac{1}{n!}\sum_{k_1\neq\dots\neq k_n=1}^n V_n(\bfr_{{k}_1},\dots,\bfr_{{k}_n})\label{eq:5}.
\end{align}
For ease of notation we also adopt the decomposition
$\bGamma_{ij}^{\mu\nu}=\gamma_{\mu\nu}\delta_{ij}\mathbf{1}+\gamma_{\mu\nu}\widetilde{\bGamma}_{ij}^{\mu\nu}$,
recalling that we are not making use of Einstein's summation convention.
Thus,
\begin{align}
 \bGamma_{ij}=\delta_{ij}\bUpsilon+\widetilde{\bUpsilon}_{ij} ,
\end{align}
with the $6\times 6$ tensor $\bUpsilon = (\gamma_{\mu\nu}\mathbf{1})$, where
$\gamma_{\mu\nu}$ is the constant friction coefficient for an isolated
particle and
$\widetilde{\bUpsilon}_{ij}=(\gamma_{\mu\nu}\widetilde{\bGamma}_{ij}^{\mu\nu})$,
where $\widetilde{\bGamma}_{ij}^{\mu\nu}$ are the translational-rotational HI
tensors. In the special case when $\gamma_{\mu\nu}=0,\ \forall\mu,\nu$, the
FPE (\ref{eq:4}) becomes the Liouville equation initially
derived by
\citeauthor{condiff_brownian_1966}\cite{condiff_brownian_1966,condiff_transport_1969,dahler_kinetic_1963}
for molecular fluids consisting of non-spherical particles. In this case one
usually invokes a Kirkwood-type time smoothing
technique\cite{condiff_brownian_1966,kirkwood_statistical_1946}, rather than
a local-equilibrium approximation involving a free-energy functional (see
Section \ref{sec:extended-unified-DDFT}).

We now define the reduced distributions obtained by integration of the phase-space PDF,
\begin{align}
 \fn(\bfr^n,\bfp^n;t)=\frac{N!}{(N-n)!} \int d\bfr^{N-n}d\bfp^{N-n}\,\fN(\bfr^N,\bfp^N;t)
\label{eq:7}
\end{align}
where $d\bfr^{N-n}\equiv d\bfr_{n+1}\dots d\bfr_N$, with an analogous expression for $d\bfp$, and the corresponding configuration-space distribution,
\begin{align}
\rhon(\bfr^n;t) = \int d\bfp^n\,\fn(\bfr^n,\bfp^n;t).
 \label{eq:8}
\end{align}
We denote the domains of integration by $\mathfrak{V}=\mathcal{V}\times\mathcal{S}$, with $\mathcal{V}=\mathbb{R}^3$ and $\mathcal{S}=[0,2\pi)\times[0,\pi)\times[0,2\pi)$, and so
$\int d\bfr \equiv \int_{\mathcal{V}}d\mathcal{V}\int_{\mathcal{S}}d\mathcal{S}$
and $\int d\bfp \equiv \int_{\mathcal{V}}d\mathcal{V}\int_{\mathcal{V}}d\mathcal{V}$, where
\begin{align}
 \int_{\mathcal{V}}d\mathcal{V}=& \int_{0}^\infty dx \int_{0}^\infty dy \int_{0}^\infty dz \label{eq:9}\\
 \int_{\mathcal{S}}d\mathcal{S}=&\int_{0}^{2\pi}d\theta \int_{0}^{\pi}d\phi\, \sin(\phi)\int_{0}^{2\pi} d\chi.
 \notag
\end{align}
For ease of notation $\rho^{(1)}$ is replaced with $\rho$, also known as the
one-body density. We now take successive momentum moments of Equation (\ref{eq:4})
starting at zeroth order and integrating over all but one-particle position
and momentum, i.e. $N \int \text{(\ref{eq:4}) } d\bfr^{N-1} d\bfp^{N}$,
resulting in the continuity equation
\begin{align}
 \partial_t\rho(\bfr_1;t) + \bnabla_{\bfr_1}\cdot\bfj(\bfr_1;t) = 0,
 \label{eq:10}
\end{align}
with the flux defined as $\bfj(\bfr_1;t)=\int d\bfp\,(\bfm^{-1}\bfp_1)f^{(1)}(\bfr_1,\bfp_1;t)$.

To obtain the first-momentum moment equation we use $N\int (\bfm^{-1}\bfp_1)\otimes \text{(\ref{eq:4}) }d\bfr^{N-1}d\bfp^N$ giving
\begin{align}
 \partial_t\bfj(\bfr_1;t)&+\bnabla_{\bfr_1}\cdot\int d\bfp_1(\bfm^{-1}\bfp_1\otimes\bfm^{-1}\bfp_1)f^{(1)}(\bfr_1,\bfp_1;t)\label{eq:11}\\
 &+\bfm^{-1}\bnabla_{\bfr_1}V_1(\bfr_1;t)\rho(\bfr_1;t)
 +\bfm^{-1}\sum_{n=2}^N\int d\bfr_2\dots d\bfr_n\bnabla_{\bfr_1}V_n(\bfr^n;t)\rho^{(n)}(\bfr^n;t)\notag\\
 &+\bUpsilon_{\parallel}\bfj(\bfr_1;t)+\bUpsilon_{\bot}\bfj_{\bot}(\bfr_1;t)+N\bfm^{-1}\sum_{j=1}^{N}\int d\bfr^{N-1}d\bfp^N\,\widetilde{\bUpsilon}_{1j}(\bfr^N)\bfp_j\fN(\bfr^N,\bfp^N;t) = 0\notag
\end{align}
where $\otimes$ is the dyadic product.  Here we have defined
$\bUpsilon_{\parallel}=\text{diag}(\gamma_{TT}\mathbf{1},\gamma_{RR}\mathbf{1})$,
$\bUpsilon_{\bot}=\text{diag}(\gamma_{TR}\mathbf{1},\gamma_{RT}\mathbf{1})$
and $\bfj_{\bot}=\int
d\bfp_1(\bfm^{-1}_{\bot}\bfp_1)f^{(1)}(\bfr_1,\bfp_1;t)$, with
$\bfm^{-1}_{\bot}=\text{adiag}(m^{-1}\mathbf{1},\mathbb{I}^{-1})$ the
anti-diagonal block matrix with submatrices, $(\bfm_{\bot}^{-1})_{1,2}=
m^{-1}\mathbf{1}$, $(\bfm_\bot^{-1})_{2,1}=\mathbb{I}^{-1}$ and
$(\bfm_\bot^{-1})_{i,j}=\mathbf{0}$ otherwise. For particles where coupling
cannot be neglected, such as screw-like
colloids\cite{wittkowski_dynamical_2011}, this first-momentum moment equation
includes an additional term compared to that for spherical particles. This
term, involving $\bfj_\bot(\bfr_1;t)$, reflects the coupled translational and
rotational forces. When such a coupling vanishes, as in the case of
orthotropic-like particles\cite{wittkowski_dynamical_2011}, the rotational
and translational motions are completely decoupled and Equation (\ref{eq:11}) can
be split into two independent equations.

Equations (\ref{eq:10})--(\ref{eq:11}), are the basis of our DDFT. In the
next section we introduce the closures which link the terms $f^{(1)}$,
$\rho^{(n)}$ and $\fN$ of Eq (\ref{eq:11}) to the one-body density, following
the same reasoning as that applied for the spherical-particle case.

\section{Extending the unified DDFT equation to non-spherical particles\label{sec:extended-unified-DDFT}}

The closure procedure will be performed in four steps: (i) the term involving
the dyadic product will be rewritten using the definition of the kinetic
stress, $\bPi(\bfr_1;t)$, and the equilibrium distribution function,
$f_{eq}^{(1)}(\bfr_1,\bfp_1)$; (ii) an adiabatic approximation will be used
to substitute the $n$-body density terms with an expression involving the
free-energy functional, $\mathcal{F}[\rho]$; (iii) the $N$-body distribution
function will be expressed as a sum of a local-equilibrium and
non-equilibrium distributions, yielding an extension of the result obtained
for spherical-colloidal fluids, and finally; (iv) we restrict our attention
to the case of two-body HIs and use an \red{Enskog-like} approximation.

\subsection{Equilibrium distribution, kinetic stress and adiabatic approximation
\label{subsec:equilibrium-kinetic-stress-adiabatic-approximation}}

At equilibrium the one-body distribution function has a Maxwell-Boltzmann structure\cite{condiff_brownian_1966}
\begin{align}
 f_{eq}^{(1)}(\bfr_1,\bfp_1)=\frac{\sqrt{|\mathbb{I}^{-1}|}}{m^{3/2}(2\pi k_BT)^3}\rho_{eq}(\bfr_1)\exp\left(-\frac{(\bfm^{-1}\bfp_1)\cdot\bfp_1}{2k_BT}\right)
 \label{eq:12},
\end{align}
with $|\mathbb{I}^{-1}|=\prod_{i}(I_{ii})^{-1}$. Hence, one can easily check that $
 \int d\bfp_1 (\bfm^{-1}\bfp_1\otimes \bfm^{-1}\bfp_1)f_{eq}^{(1)}(\bfr_1,\bfp_1)=  k_BT\bfm^{-1}\rho_{eq}(\bfr_1)$.
By adding and subtracting $\bnabla_{\bfr_1}\cdot \left(
k_BT\bfm^{-1}\rho(\bfr_1)\right)$ on the left-hand side of Equation (\ref{eq:11})
we can write,
\begin{align}
 k_BT\bfm^{-1}\bnabla_{\bfr_1}\rho(\bfr_1;t)+\bnabla_{\bfr_1}\cdot\bPi(\bfr_1;t)&=\,\bnabla_{\bfr_1}\cdot\int d\bfp_1(\bfm^{-1}\bfp_1\otimes\bfm^{-1}\bfp_1)f^{(1)}(\bfr_1,\bfp_1;t) \label{eq:13}
\end{align}
with the kinetic-stress tensor defined as\cite{goddard_unification_2013}
\begin{align}
 \bPi(\bfr_1;t)&=\,\int d\bfp_1\left((\bfm^{-1}\bfp_1\otimes\bfm^{-1}\bfp_1)-k_BT\bfm^{-1}\right)f^{(1)}(\bfr_1,\bfp_1;t)\label{eq:14}.
\end{align}
According to this definition, the term involving the one-body distribution
function and the dyadic product in
Equation (\ref{eq:11}) can be
identified as the remainder of the spatial and angular divergence of the
kinetic stress. Although at this stage Equation (\ref{eq:13}) only seems to be a
more convenient way of expressing the second term on the left-hand side of Equation (\ref{eq:11}),
the advantage will become clear in Section
\ref{subsec:local-equilibrium} where we discuss how to deal with
out-of-equilibrium distributions.

But before this, we will make use of the adiabatic approximation. This
approximation identifies, at any time $t$, the one-body density,
$\rho(\bfr_1;t)$, with the equilibrium one, $\rho_{eq}(\bfr_1)$. We now
consider the Helmholtz free-energy
functional\cite{evans_nature_1979,lutsko_recent_2010}
\begin{align}
 \mathcal{F}[\rho_{eq}]=\red{k_BT}\int_{\mathfrak{V}}d\bfr_1\,\rho_{eq}(\bfr_1)\left(\ln\left(\Lambda^{3}\rho_{eq}(\bfr_1)-1\right)\right)
 +\mathcal{F}_{exc}[\rho_{eq}]+\int_{\mathfrak{V}}d\bfr_1\,\rho_{eq}(\bfr_1)V_1(\bfr_1)
 \label{eq:15}
\end{align}
where the first term on the right-hand side accounts for the ideal-rotator
gas component, the last term contains the external potential contribution to
the free energy, $\mathcal{F}_{exc}$ is an excess over ideal gas contribution
and $\Lambda$ is the de Broglie wavelength (which will turn out to be
irrelevant). From classical DFT\cite{lutsko_recent_2010} we know that there
is a unique external potential, $\mathfrak{W}(\bfr_1)=w(\bfr_1)+V_1(\bfr_1)$,
for which $\rho_{eq}(\bfr_1)$ is the equilibrium density. \red{Following
\citet{rex_lowen_dynamical_2009}, the following generalised force-balance
equation is fulfilled,}
\begin{equation}
 \red{
 \frac{k_BT\bnabla_{\bfr_1}\rho(\bfr_1)}{\rho(\bfr_1)}+\bnabla_{\bfr_1}\mathfrak{W}(\bfr_1) =
 -\left.\bnabla_{\bfr_1}\frac{\delta \mathcal{F}_{exc}[\rho]}{\delta \rho(\bfr_1)}\right|_{\rho=\rho_{eq}}}
 \label{eq:16-a}
\end{equation}
\red{which makes it possible to determine the gradient of the additional
potential $w(\bfr_1)$ introduced in the definition of the external potential.
Rearranging Eq. (\ref{eq:16-a}),}
\begin{equation}
 \red{\bnabla_{\bfr_1}w(\bfr_1)=-\left.\bnabla_{\bfr_1}\frac{\delta \mathcal{F}[\rho]}{\delta \rho(\bfr_1)}\right|_{\rho=\rho_{eq}}\equiv
 -\bnabla_{\bfr_1}\mu(\bfr_1)
 }
 \label{eq:16}
\end{equation}
\red{with $\mu(\bfr)$ the local nonequilibrium chemical potential. This
way, $\bnabla_{\bfr_1}w(\bfr_1)$ can be understood as the additional external
force the system needs to balance the thermodynamic driving
force\cite{rex_lowen_dynamical_2009}. }

Furthermore, under the same circumstances Equation (\ref{eq:11}) becomes the first equation of Yvon-Born-Green (YBG) hierarchy\cite{barrat_basic_2003,liboff_kinetic_2003}
\begin{equation}
 k_BT\bfm^{-1}\bnabla_{\bfr_1}\rho_{eq}(\bfr_1)+\bfm^{-1}\bnabla_{\bfr_1}\mathfrak{W}(\bfr_1;t)\rho_{eq}(\bfr_1)=
 \bfm^{-1}\sum_{n=2}^N\int d\bfr_2\dots d\bfr_n\bnabla_{\bfr_1}V_n(\bfr^n)\rho^{(n)}_{eq}(\bfr^n)
 \label{eq:17}
\end{equation}
Making the approximation that Equations (\ref{eq:15})--(\ref{eq:17}) hold out of
equilibrium, i.e. substitute $\rho_{eq}^{(n)}$ by $\rho^{(n)}(t)$ for all
$n$, yields,
\begin{align}
 \rho(\bfr_1;t)\bnabla_{\bfr_1}\frac{\delta \mathcal{F}[\rho]}{\delta \rho(\bfr_1)}=&\,
 \rho(\bfr_1;t)\bnabla_{\bfr_1}V_1(\bfr_1;t)+k_BT\bnabla_{\bfr_1}\rho(\bfr_1;t)\label{eq:18}\\
 &+\sum_{n=2}^N\int d\bfr_2\dots d\bfr_n\bnabla_{\bfr_1}V_n(\bfr^n)\rho^{(n)}(\bfr^n;t). \notag
\end{align}
Substituting Equations (\ref{eq:13}) and (\ref{eq:18}) into Equation (\ref{eq:11}) produces,
\begin{align}
 \partial_t\bfj(\bfr_1;t)+&\bnabla_{\bfr_1}\cdot\bPi(\bfr_1;t)+\bfm^{-1}\rho(\bfr_1;t)\bnabla_{\bfr_1}\frac{\delta \mathcal{F}[\rho]}{\delta \rho(\bfr_1)}
 +\bUpsilon_{\parallel}\bfj(\bfr_1;t)+\bUpsilon_{\bot}\bfj_{\bot}(\bfr_1;t)\label{eq:19}\\
 +&N\bfm^{-1}\sum_{j=1}^{N}\int d\bfr^{N-1}d\bfp^N\,\widetilde{\bUpsilon}_{1j}(\bfr^N)\bfp_j\fN(\bfr^N,\bfp^N;t) = 0.\notag
\end{align}

\subsection{Local-equilibrium approximation and beyond\label{subsec:local-equilibrium}}

The next step of the derivation is the introduction of a widely used
approximation in statistical
physics\cite{pottier_nonequilibrium_2014,liboff_kinetic_2003,goddard_unification_2013},
namely the local-equilibrium approximation. It allows rewriting the kinetic
stress in terms of the one-body density. With the definitions of local
density, $\rho$, velocity, $\bfv$, and temperature, $T$, given by
\begin{align}
    \int d\bfp \left\{\begin{array}{c}
             1\\
             \bfp\\
             (\bfm^{-1}\mathbf{c})\cdot\mathbf{c}
            \end{array}\right\}
	      f^{(1)}(\bfr,\bfp;t)=
	      \left\{\begin{array}{c}
	       1\\
	       \bfm\bfv(\bfr;t)\\
	       6k_BT
	      \end{array}\right\}\rho(\bfr;t)
	      \label{eq:20},
\end{align}
with\cite{liboff_kinetic_2003}
$\mathbf{c}\equiv\mathbf{c}(\bfr,\bfp;t)=\bfp-\bfm\bfv(\bfr;t)$ the
deviation-from-the-mean microscopic momentum, the local-equilibrium
approximation to the true probability distribution,
\begin{align}
 f^{(1)}_{leq}(\bfr,\bfp;t)=&\, \frac{|\mathbb{I}^{-1}|^{1/2}\rho(\bfr;t)}{m^{3/2}(2\pi k_BT)^{3}}\label{eq:21}
  \exp\left(-\frac{(\bfm^{-1}\mathbf{c})\cdot\mathbf{c}}{2k_BT}\right)
\end{align}
fulfils the normalisation conditions given in Equation (\ref{eq:20}). While
approximating $f^{(1)}$ with Equation (\ref{eq:21}) can be reasonable as a
first-order approach, there is in fact no actual reason to impose such a
restriction. We instead introduce the expansion
$f^{(1)}=f^{(1)}_{leq}\,\left[1+\varPsi \right]\equiv
f^{(1)}_{leq}+f^{(1)}_{neq}$, which takes into account both local-equilibrium
and non-equilibrium effects.  The non-equilibrium term must satisfy
\begin{align}
 \int d\bfp \left\{\begin{array}{c}
             1\\
             \bfp\\
             (\bfm^{-1}\mathbf{c})\cdot\mathbf{c}
            \end{array}\right\}
	      f^{(1)}_{neq}(\bfr,\bfp;t)=
	      \left\{\begin{array}{c}
	       0\\
	       \mathbf{0}\\
	       0
	      \end{array}\right\}.\label{eq:22}
\end{align}
Hence, $f_{neq}^{(1)}$ only contributes to higher moments such as the stress,
$\bPi$, and the heat flux, $\boldsymbol{\mathfrak{Q}}$. Using Equations
(\ref{eq:20})--(\ref{eq:22}) it is readily found that
$\bfj(\bfr;t)=\bfj_{leq}(\bfr;t)=\rho(\bfr;t)\bfv(\bfr;t)$. Thus, recalling
the definition given in Equation (\ref{eq:11}), the kinetic-stress tensor
satisfies the decomposition,
\begin{align}
 \bPi(\bfr;t)=& (\bfv(\bfr;t)\otimes\bfv(\bfr;t))\rho(\bfr;t)+
 \int d \bfp\left(\bfm^{-1}\bfp\otimes \bfm^{-1}\bfp\right)f_{neq}^{(1)}(\bfr,\bfp;t)\label{eq:23} =\bPi_{leq}(\bfr;t) + \bPi_{neq}(\bfr;t),
\end{align}
where the first term stands for the local-equilibrium stress, which embodies
the kinetic contribution\cite{condiff_transport_1969} due to translational,
rotational and coupled velocities. On the other hand, the heat flux only
shows a non-equilibrium contribution and, given the symmetry of
$f_{leq}^{(1)}$,
\begin{align}
 \boldsymbol{\mathfrak{Q}}(\bfr;t)=\frac{1}{2}\int d\bfp\,|\mathbf{c}|^2\,\bfm^{-1}\mathbf{c}\,f^{(1)}(\bfr,\bfp;t)
=\frac{1}{2}\int d\bfp\,|\mathbf{c}|^2\,\bfm^{-1}\mathbf{c}\,f_{neq}^{(1)}(\bfr,\bfp;t)\label{eq:24}.
 \end{align}
Now, making use of the identity
\begin{align}
\bnabla_{\bfr}\cdot\left[(\bfv(\bfr;t)\otimes\bfv(\bfr;t))\rho(\bfr;t)\right] =
\bfv(\bfr;t)\bnabla_{\bfr}\cdot\left(\bfv(\bfr;t)\rho(\bfr;t)\right)+\rho(\bfr;t)\left(\bfv(\bfr;t)\cdot\bnabla_{\bfr}\right)\bfv(\bfr;t),\label{eq:25}
\end{align}
in Equations (\ref{eq:23}) and (\ref{eq:19}), we arrive at the time-evolution equation
\begin{align}
 \rho(\bfr_1;t)\mathcal{D}_t\bfv(\bfr_1;t)\label{eq:26}+&\bnabla_{\bfr_1}\cdot \bPi_{neq}(\bfr_1;t)+\bfm^{-1}\rho(\bfr_1;t)\bnabla_{\bfr_1}\frac{\delta \mathcal{F}[\rho]}{\delta \rho(\bfr_1)}
 +(\bUpsilon_{\parallel}+\bUpsilon_{\bot}\bepsilon)\rho(\bfr_1;t)\bfv(\bfr_1;t)\\
 +&N\bfm^{-1}\sum_{j=1}^{N}\int d\bfr^{N-1}d\bfp^N\,\widetilde{\bUpsilon}_{1j}(\bfr^N)\bfp_j\fN(\bfr^N,\bfp^N;t) = 0\notag
\end{align}
where we define the material derivative as
$\mathcal{D}_t=(\partial_t+\bfv(\bfr;t)\cdot\bnabla_{\bfr})$ and
$\bepsilon=\text{adiag}(m^{-1}\mathbb{I},m\mathbb{I}^{-1})$. At this stage,
there are two terms which still have to be treated, namely the
non-equilibrium and the friction tensor terms. The former will be analysed in
what follows and the latter in Section\ \ref{subsec:enskog-closure}.

To treat the non-equilibrium term, we follow Grad's
method\cite{grad_kinetic_1949,liboff_kinetic_2003}. Specifically, the
one-body distribution is rewritten as $f^{(1)}=f^{(1)}_{leq}\,\left[1+\varPsi
\right]$ and $\Psi(\bfr,\bfp;t)$ is expanded in a basis of orthogonal
polynomials of the $\mathbf{p}\oplus\mathbf{L}$ space. In particular, we
choose a basis of the multiple generalised Hermite
polynomials\cite{dunkl_classical_2001} on $\mathbb{R}^d$,
\begin{align}
P^{[\alpha]}_{n,\boldsymbol{\ell}}(x_1,\dots,x_d)=H_{\ell_1}^{[\alpha]}(x_1)\dots H_{\ell_d}^{[\alpha]}(x_d)\label{eq:27},
\quad \sum\ell_k=n
\end{align}
with $n\in\mathbb{N}_0=\mathbb{Z}^+\cup\{0\}$,
$\boldsymbol{\ell}\in\mathbb{N}^d_0$ and $H_{k}^{[\alpha]}(x)\equiv
H_{k}(x/\sqrt{\alpha})/\sqrt{k!}$ the generalised Hermite polynomials, which
are an orthogonal basis for $L^2\left(\mathbb{R}\right)$ with respect to the
weight $(2\pi\alpha)^{-1/2} \exp( -x^2/(2\alpha) )$. Denoting
$\mathbf{e}_j=(\delta_{jk})$, the first
$P^{[\alpha]}_{n,\boldsymbol{\ell}}$\, are given by
\begin{align}
 P^{[\alpha]}_{0,\boldsymbol{0}}(\mathbf{x})=\,1,\label{eq:28}\quad P^{[\alpha]}_{1,\mathbf{e}_{j}}(\mathbf{x})=\,\frac{x_j}{\sqrt{\alpha}},\quad
 P^{[\alpha]}_{2,\mathbf{e}_{j}+\mathbf{e}_{k}}(\mathbf{x})=\,\frac{x_jx_k}{\alpha}-\delta_{jk}.
\end{align}
Setting now $\alpha= \sqrt{k_BT}$, the one-body distribution function can be
expanded as
\begin{align}
 f^{(1)}(\bfr,\bfp;t)=f^{(1)}_{leq}(\bfr,\bfp;t)\left(1+\sum_{n,\boldsymbol{\ell}}B_{n,\boldsymbol{\ell}}(\bfr;t)\,
 P^{[\alpha]}_{n,\boldsymbol{\ell}}\left(\bfm^{-1/2}\mathbf{c}\right)\right).\label{eq:29}
\end{align}
with $B_{n,\boldsymbol{\ell}}(\bfr;t)$ the moments of the distribution
$f_{neq}^{(1)}$. From Equation (\ref{eq:22}), we then find that
$B_{0,\mathbf{0}}=B_{1,\mathbf{e}_j}=0$ and
$\sum_{j}{B}_{2,2\mathbf{e}_j}=0$. Finally, the non-equilibrium term in Equation
(\ref{eq:26}) can be expressed as
\begin{align}
\boldsymbol{\Pi}_{jk}^{neq}(\bfr;t) = k_B T \rho(\bfr;t) B_{2,\mathbf{e}_j+\mathbf{e}_k}(\bfr;t),
\label{eq:30}
\end{align}
which represents an extension of the analogous relationship found for the spherical case (Section 3.2 of Ref. \onlinecite{goddard_unification_2013}). This relationship does not necessarily imply the non-local kinetic pressure to be small, since the functions $B_{2,\mathbf{e}_j+\mathbf{e}_k}(\bfr;t)$ do not depend on $\rho,\, T$ or $\bfv$.
However,
at this point we make the assumption that $f_{neq}^{(1)}$  can be either
approximated or neglected,
inspired by previous rigorous analyses for spherical
particles\cite{goddard_overdamped_2012,goddard_unification_2013}. Therefore,
to obtain a DDFT, we need to deal with the last term in Equation
(\ref{eq:26}).

\subsection{Two-body interactions and closure equation\label{subsec:enskog-closure}}

With the aim of closing Equation (\ref{eq:26}) in $\rho$, we restrict our
attention to systems where the HIs are given by linear combinations of
two-body interactions. In particular, we assume the general
structure\cite{goddard_unification_2013,goddard_overdamped_2012}
\begin{align}
 \widetilde{\bUpsilon}_{jk}(\bfr^N)=\delta_{jk}\sum_{r\neq j}\mathbf{Z}_1(\bfr_j,\bfr_r)+(1-\delta_{jk})\mathbf{Z}_2(\bfr_j,\bfr_k)\label{eq:31}.
\end{align}
The integral of Equation (\ref{eq:26}) depends only on
$f^{(2)}(\bfr_1,\bfr_2,\bfp_1,\bfp_2;t)$ and we now need a closure relation
between the one- and two-body distributions. From a previous rigorous derivation of
DDFT\cite{chan_time-dependent_2005},
it is known that the one-body density determines the $N$-body distribution
and, therefore, all the time-dependent properties of the system can be
expressed as functionals of $\rho(\bfr;t)$. Thus, we make the \red{following}
approximation
\begin{align}
f^{(2)}(\bfr_1,\bfr_2,\bfp_1,\bfp_2;t) = f^{(1)}(\bfr_1,\bfp_1;t)f^{(1)}(\bfr_2,\bfp_2;t)g(\bfr_1,\bfr_2;[\rho]),
\label{eq:32}
\end{align}
although the exact form of the functional $g(\bfr_1,\bfr_2;[\rho])$ is
unknown, accurate approximations are known at equilibrium.
However, the problem of obtaining an expression for $g$ is beyond the scope
of the present study. From now on, we assume there exists at least a good
approximation. \red{It is also worth mentioning that
despite being analogous in structure to the Enskog approximation, Eq. (\ref{eq:32}) should be only understood as a local-equilibrium approximation. Thanks to this last closure, we no longer have to deal with higher order distributions but only with the one-body one.}

Substituting Equations (\ref{eq:31}) and (\ref{eq:32}) into (\ref{eq:30}) yields
\begin{align}
 \rho(\bfr_1;t)\mathcal{D}_t\bfv(\bfr_1;t)+&\bnabla_{\bfr_1}\cdot \bPi_{neq}(\bfr_1;t)+\bfm^{-1}\rho(\bfr_1;t)\bnabla_{\bfr_1}\frac{\delta \mathcal{F}[\rho]}{\delta \rho(\bfr_1)}
 +(\bUpsilon_{\parallel}+\bUpsilon_{\bot}\bepsilon)\rho(\bfr_1;t)\bfv(\bfr_1;t)\label{eq:33}\\
+&\bfm^{-1}\rho(\bfr_1;t)\int d\bfr_2\left[
 \mathbf{Z}_1(\bfr_1,\bfr_2)\bfm\bfv(\bfr_1;t)+\mathbf{Z}_2(\bfr_1,\bfr_2)\bfm\bfv(\bfr_2;t)
 \right]\,\rho(\bfr_2;t)\,g(\bfr_1,\bfr_2;[\rho]) = 0. \notag
\end{align}
Neglecting the non-local equilibrium term as discussed above, and dividing by
the one-body density, we finally obtain our equation for the evolution of the
flux,
\begin{align}
 \mathcal{D}_t\bfv(\bfr_1;t)+&\bfm^{-1}\bnabla_{\bfr_1}\frac{\delta \mathcal{F}[\rho]}{\delta \rho(\bfr_1)}\label{eq:34}
 +(\bUpsilon_{\parallel}+\bUpsilon_{\bot}\bepsilon)\bfv(\bfr_1;t)\\
 +&\bfm^{-1}\int d\bfr_2\left[
 \mathbf{Z}_1(\bfr_1,\bfr_2)\bfm\bfv(\bfr_1;t)+\mathbf{Z}_2(\bfr_1,\bfr_2)\bfm\bfv(\bfr_2;t)
 \right]\,\rho(\bfr_2;t)\,g(\bfr_1,\bfr_2;[\rho]) = 0\notag.
\end{align}
Combining this equation with the conservation law resulting from Equations
(\ref{eq:20}) and (\ref{eq:22}) in (\ref{eq:10}), namely
\begin{align}
 \partial_t\rho(\bfr_1;t)+\bnabla_{\bfr_1}\cdot(\rho(\bfr;t)\bfv(\bfr_1;t))=0\label{eq:35},
\end{align}
we finally have the generalised DDFT for systems of orientable particles.

We can readily check that for the special cases of point-like and smooth
spherical particles (with perfect slip on the
surface\cite{dickinson_brownian_1985}), these equations reduce to the unified
DDFT recently derived by \citet{goddard_unification_2013}. Section \ref{subsec:wl-eq}
will be devoted to showing how Equation (\ref{eq:34}) turns into the
equation derived by \citet{rex_dynamical_2007} and
\citet{wittkowski_dynamical_2011}
for systems of arbitrary-shape colloids by setting
$\mathbf{Z}_1=\mathbf{Z}_2=0$, and showing that the generalised Einstein
relationships\cite{dickinson_brownian_1985,peters_smoluchowski_2000} can be
recovered in such a case. Finally, we note that the overall effect of the last
two terms in Equation (\ref{eq:34}) will be a friction-like retardation of the
translational and angular velocities $\bfv(\bfr_1,t)$, akin to what happened for spherical
particles\cite{goddard_unification_2013}.

%

\subsection{Connection with previous DDFTs\label{subsec:wl-eq}}

Here we connect our DDFT with the DDFTs obtained by
\citet{rex_dynamical_2007} and \citet{wittkowski_dynamical_2011} for systems
of arbitrary-shape colloids with overdamped dynamics, neglecting HIs. Within
the context of our study, this is equivalent to setting
$\mathbf{Z}_1=\mathbf{Z}_2=0$. We then have
\begin{align}
  (\bUpsilon_{\parallel}+\bUpsilon_{\bot}\bepsilon)\bfv(\bfr_1;t) = -\mathcal{D}_t\bfv(\bfr_1;t)-&\bfm^{-1}\bnabla_{\bfr_1}\frac{\delta \mathcal{F}[\rho]}{\delta \rho(\bfr_1)}\label{eq:36}.
\end{align}
Using 
this result in the continuity equation (\ref{eq:35}) gives
\begin{align}
 \partial_t\rho(\bfr_1;t)=\bnabla_{\bfr_1}\cdot\left(\rho(\bfr_1;t)\boldsymbol{\zeta}^{-1}\left(\bnabla_{\bfr_1}\frac{\delta \mathcal{F}[\rho]}{\delta \rho(\bfr_1)}+\bfm\mathcal{D}_t\bfv(\bfr_1;t)\right)\right)\label{eq:37}
\end{align}
where we defined
 \begin{align}
(\bUpsilon_{\parallel}+\bUpsilon_{\bot}\bepsilon)\bfm=\boldsymbol{\zeta}\equiv\,\begin{pmatrix}
                                                           \boldsymbol{\zeta}_{\text{TT}} & \boldsymbol{\zeta}_{\text{TR}}\\
                                                           \boldsymbol{\zeta}_{\text{RT}} & \boldsymbol{\zeta}_{\text{RR}}
                                                          \end{pmatrix}\label{eq:38}
 \end{align}
along with  $\boldsymbol{\zeta}_{\text{TT}} =\, m\gamma_{TT}\mathbf{1},\ \boldsymbol{\zeta}_{\text{TR}} =\, m^{-1}\gamma_{TR}\mathbb{I}^2,\ \boldsymbol{\zeta}_{\text{RT}} =\, m^2\gamma_{RT}\mathbb{I}^{-1},$ and $\boldsymbol{\zeta}_{\text{RR}} =\, \gamma_{RR}\mathbb{I}$. In the overdamped regime, inertial forces and unsteady acceleration are neglected, i.e. $\bfm\,\mathcal{D}_t\bfv\rightarrow0$. Therefore, Equation (\ref{eq:38}) becomes
\begin{align}
\partial_t\rho(\bfr_1;t)=\bnabla_{\bfr_1}\cdot\left(\rho(\bfr_1;t)\beta\mathbf{D}(\bfr_1;t)\left(\bnabla_{\bfr_1}\frac{\delta \mathcal{F}[\rho]}{\delta \rho(\bfr_1)}\right)\right)\label{eq:39}
\end{align}
where $\beta\mathbf{D}=\boldsymbol{\zeta}^{-1}$ is the diffusion tensor.
Making use of the block-matrix inversion formula [e.g. Equation (2.8.18) of
Ref. \onlinecite{bernstein_matrix_2005}] we can check that the diffusion tensor fulfils
the generalised Einstein
relationships\cite{dickinson_brownian_1985,peters_smoluchowski_2000}
 \begin{align}
  \beta\mathbf{D}_{\text{TT}} =& \left(\boldsymbol{\zeta}_{\text{TT}}-\boldsymbol{\zeta}_{\text{TR}}\boldsymbol{\zeta}_{\text{RR}}^{-1}\boldsymbol{\zeta}_{\text{RT}}\right)^{-1}\label{eq:40}\\
  \beta\mathbf{D}_{\text{TR}} =& -\left(\boldsymbol{\zeta}_{\text{TT}}-\boldsymbol{\zeta}_{\text{TR}}\boldsymbol{\zeta}_{\text{RR}}^{-1}\boldsymbol{\zeta}_{\text{RT}}\right)^{-1}\boldsymbol{\zeta}_{\text{TR}}\boldsymbol{\zeta}_{\text{RR}}^{-1}\equiv
  \left(\boldsymbol{\zeta}_{\text{RT}}-\boldsymbol{\zeta}_{\text{TT}}\boldsymbol{\zeta}_{\text{TR}}^{-1}\boldsymbol{\zeta}_{\text{RR}}\right)^{-1}\notag\\
  \beta\mathbf{D}_{\text{RT}} =& -\left(\boldsymbol{\zeta}_{\text{TT}}-\boldsymbol{\zeta}_{\text{TR}}\boldsymbol{\zeta}_{\text{RR}}^{-1}\boldsymbol{\zeta}_{\text{RT}}\right)^{-1}\boldsymbol{\zeta}_{\text{RT}}\boldsymbol{\zeta}_{\text{TT}}^{-1}
  \equiv\left(\boldsymbol{\zeta}_{\text{TR}}-\boldsymbol{\zeta}_{\text{RR}}\boldsymbol{\zeta}_{\text{RT}}^{-1}
  \boldsymbol{\zeta}_{\text{TT}}\right)^{-1}\notag\\
  \beta\mathbf{D}_{\text{RR}} =& \left(\boldsymbol{\zeta}_{\text{RR}}-\boldsymbol{\zeta}_{\text{RT}}\boldsymbol{\zeta}_{\text{TT}}^{-1}\boldsymbol{\zeta}_{\text{TR}}\right)^{-1}\notag
 \end{align}
In the special case when the translational and rotational motions are fully
decoupled, $\mathbf{D}_{\text{TR}}=\mathbf{D}_{\text{RT}}\equiv 0$, we
recover the equation derived by \citet{rex_dynamical_2007}. This is indeed
the case for spherically isotropic colloids [e.g. Equation (37) of
Ref. \onlinecite{dickinson_brownian_1985}] far from a wall.
Furthermore, if the colloidal particles are smooth spheres we recover the
DDFT derived by \citet{marconi_dynamic_2000}, as the stress tensor is then
uniquely determined by the translational component. However, the strength of
the hydrodynamic translation-rotation coupling increases rapidly as the
distance of the particles from a wall
decreases\cite{dickinson_brownian_1985,beenakker_many-sphere_1984}. The
coupling is then unavoidable near the walls even for the simple spherical
case. In this regard, Equation (\ref{eq:39}) can be seen as a generalisation of
the uncoupled DDFT obtained by \citet{rex_dynamical_2007}.

Of course, our overdamped DDFT (\ref{eq:39}) does not include the
self-propulsion forces present in the DDFT of
\citet{wittkowski_dynamical_2011} as these forces were not present in our original
equations to begin with (see Appendix \ref{app:projection-operator-tech}).
Nevertheless, they can readily be included as an additional term,
$\bfF_{i}^{A}(\bfr_i;t)$, inside the solvent-averaged interaction forces and
torques $\bfF$. This results in Equation (\ref{eq:11}) having the extra term,
$-\bfm^{-1}\bfF_{1}^{A}(\bfr_1;t)\rho(\bfr_1;t)$. Following exactly the same
steps as in (\ref{eq:12})--(\ref{eq:16}), the same first equation of the YBG
hierarchy (\ref{eq:17}) can be obtained, by setting $\bfF_1^A=0$ at
equilibrium\cite{wittkowski_dynamical_2011}. Therefore, the additional term
$-\bfm^{-1}\bfF_{1}^{A}(\bfr_1;t)\rho(\bfr_1;t)$ would survive throughout
the whole derivation so that the left-hand side of our DDFT equation
(\ref{eq:34}) would contain it. Eventually, this leads us to
\begin{align}
\partial_t\rho(\bfr_1;t)=\bnabla_{\bfr_1}\cdot\left(\rho(\bfr_1;t)\beta\mathbf{D}(\bfr_1;t)\left(\bnabla_{\bfr_1}\frac{\delta \mathcal{F}[\rho]}{\delta \rho(\bfr_1)}-\bfF_{1}^{A}(\bfr_1;t)\right)\right)\label{eq:41}
\end{align}
which is now in complete agreement with the equation previously derived
within the overdamped regime\cite{wittkowski_dynamical_2011}, with the
diffusion tensor still satisfying the generalised Einstein relationships
(\ref{eq:40}).

\section{Conclusions}
\label{sec:conclusions}

In this work we have formulated a DDFT for orientable particles with inertia
and HIs.
There are numerous examples where orientation of particles plays a key role.
From the fundamental study of perfectly-rough
spheres\cite{condiff_brownian_1966}, loaded
spherocylinders\cite{condiff_brownian_1966,dahler_kinetic_1963}, nematic
solutions\cite{hartel_towing_2010,zhang_isotropic-nematic_2006} or
liquid-crystal nucleation\cite{schilling_self-poisoning_2004,
zhang_isotropic-nematic_2006, miller_hierarchical_2009}, to the study of
biological processes such as\cite{dickinson_brownian_1985} protein adsorption
and trapping, antibody-antigen interaction, biochemical assembly by monomer
aggregation or polymerization, bone
formation\cite{cantaert_nanoscale_2013,gomez-morales_progress_2013} or
\emph{in vivo} protein
crystallization\cite{bechtel_electron_1976,koopmann_vivo_2012}, to name but a
few. We believe that all these problems could be tackled with the aid of the
DDFT developed here.

The present study also addresses the problem posed by
\citet{wittkowski_dynamical_2011}: generalising their work to include HIs.
But our study goes a step further and also includes the inertia of the
colloidal particles. We are currently in the process of implementing our DDFT
equations in the pseudospectral scheme, based on Chebychev
collocation\cite{boyd_chebyshev_2001}, we have developed for the
numerical treatment of DDFT for spherical particles and which has been
successfully applied to a wide spectrum of physical settings; from fluids in
a confining potential\cite{goddard_unification_2013} and mixtures in such a
potential\cite{goddard_multi-species_2013} to adsorbed films on a
substrate\cite{yatsyshin_spectral_2012}, fluids in confined
geometry\cite{yatsyshin_geometry-induced_2013,yatsyshin_wetting_2015,yatsyshin_density_2015} and contact
lines\cite{nold_fluid_2014,nold_nanoscale_2015}. While the theory presented here is
valid for three-dimensional orientable particles, from a computational point
of view systems of two-dimensional orientable particles would be more
tractable. Finally, interesting extensions of the framework developed here
would be to include torsional degrees of freedom to describe systems of
flexible-chain molecules\cite{evans_momentum_1980}, consider
multiple-particle species, thus extending the previously developed DDFT for
mixtures of spherical particles\cite{goddard_multi-species_2013}, and
confined geometry\cite{yatsyshin_geometry-induced_2013,yatsyshin_wetting_2015,yatsyshin_density_2015}. We shall
examine these and related questions in future studies.

\begin{acknowledgments}
We are grateful to the anonymous referees for useful comments and suggestions
and to Andreas Nold for stimulating discussions that led to the scaling
arguments in Appendix B. We acknowledge financial support from the European
Research Council via Advanced Grant No. 247031 and from EPSRC via grants No.
EP/L020564 and EP/L025159.
\end{acknowledgments}

\appendix
\numberwithin{equation}{section}

\section{Derivation of generalised Langevin equations}\label{app:projection-operator-tech}

\subsection{The Fokker-Planck Equation}
Here we outline the derivation of the time-evolution equation for the
probability distribution function of a system of arbitrary-shape particles.
This derivation can be thought as an application of the works of
\citet{kirkwood_statistical_1946}, \citet{murphy_brownian_1972},
\citet{deutch_molecular_1971} and \citet{wilemski_derivation_1976} in
conjunction with the results of \citet{dahler_kinetic_1963},
\citet{condiff_brownian_1966}, \citet{evans_cumulant_1976} and
\citet{condiff_transport_1969}, to consider arbitrary-shape particles.
\red{The derivation presented below can be also understood as a generalisation of the one carried by \citet{archer_dynamical_2005}
within the context of point-like particles}.

As in Section \ref{sec:evolution-equations}, consider $N$ identical, asymmetric colloidal particles with mass $m$
immersed in a fluid of $n\gg N$ identical bath particles with mass $m_b$.
Throughout this section upper case
letters will refer to colloidal particles while lower case ones indicate bath particles.
The phase-space coordinates of the $i$-th bath particle are $\mathbf{x}_j\doteq(\mathbf{r}_j,\,\mathbf{p}_j)$,
with ``$\doteq$'' denoting ``by definition", $\mathbf{r}_j$ being the
position vector and $\mathbf{p}_j=m_b\dot{\mathbf{r}}_j$ the canonical
momentum. In addition, the dynamical state of the $j$-th colloidal particle
is determined by $\mathbf{X}_j\doteq(\mathbf{R}_j,\mathbf{P}_j)$,  where
$\mathbf{R}_j$ is its centre-mass position vector and
$\mathbf{P}_j=m\dot{\mathbf{R}}_j$ is its conjugate momentum, along with the
pair $\boldsymbol{\Omega}_j\doteq(\boldsymbol{\alpha}_j,\boldsymbol{\pi}_j)$
which comprises the rotational degrees of freedom, with
$\boldsymbol{\alpha}_j$ being the Eulerian angles as in Section
\ref{sec:evolution-equations} and $\boldsymbol{\pi}_j$ their conjugate
momenta.\cite{dahler_kinetic_1963} The angular velocity of a particle,
$\boldsymbol{\omega}_j$, is determined by the time derivative of the Euler
angles. In the principal-axes frame $\mathfrak{B}$,
we have the relation
$\boldsymbol{\omega}_j=\boldsymbol{\Lambda}_j^\top\dot{\boldsymbol{\alpha}}_j$,
where\cite{jose_classical_2013,condiff_brownian_1966}
\begin{equation}
 \boldsymbol{\Lambda}_j^\top = \begin{pmatrix}
  \cos\chi_j & \sin\theta_j\sin\chi_j & 0\\
  -\sin\chi_j & \sin\theta_j\cos\chi_j & 0\\
  0 & \cos\theta_j & 1
 \end{pmatrix}\ \Leftrightarrow
 \boldsymbol{\Lambda}_j^{-1} =
 \begin{pmatrix}
  \cos\chi_j & \csc\theta_j\cos\chi_j & -\cot\theta_j\sin\chi_j\\
  -\sin\chi_j & \csc\theta_j\sin\chi_j & -\cot\theta_j\cos\chi_j\\
  0 & 0 & 1
 \end{pmatrix}.
 \label{eq:a-1}
\end{equation}
Accordingly, $\boldsymbol{\omega}_j'=\mathcal{R}_j^\top\boldsymbol{\omega}_j=
\boldsymbol{\Xi}_j\,\dot{\boldsymbol{\alpha}}_j$ under the space-fixed frame
$\mathfrak{S}$, with
$\boldsymbol{\Xi}_j\doteq\mathcal{R}_j^\top\boldsymbol{\Lambda}_j^\top$.
These can be related with their corresponding angular momenta by,
$\mathbf{L}_j=\mathbb{I}\,\boldsymbol{\omega}_j$ and
$\mathbf{L}'_j=\mathbb{I}'_j\boldsymbol{\omega}'_j$ respectively.

Thus, the dynamical state of the system  at any given instant represents a single point in a $6(N+n)-$dimensional space, $\Gamma$,\cite{liboff_kinetic_2003}
\begin{equation}
 \mathfrak{s}(t)\doteq(\mathbf{x}_1(t),\dots,\mathbf{x}_n(t),\mathbf{X}_1(t),\dots,\mathbf{X}_N(t),\boldsymbol{\Omega}_1(t),\dots,\boldsymbol{\Omega}_N(t))\equiv
 (\mathbf{x}^n(t),\mathbf{X}^N(t),\boldsymbol{\Omega}^N(t))\in\Gamma
 \label{eq:a-2}
\end{equation}
where we made use of the notation $\mathbf{x}^n= \mathbf{x}_1\dots
\mathbf{x}_n$, $\mathbf{X}^N=\mathbf{X}_1\dots\mathbf{X}_N$ and
$\boldsymbol{\Omega}^N=\boldsymbol{\Omega}_1,\dots,\boldsymbol{\Omega}_N$.
From classical mechanics, the evolution of the system is completely
determined by the initial conditions for positions and momenta of all
particles. This time evolution, which prescribes a unique trajectory
$\mathfrak{s}(t;t_0)$, is fully described by the Lagrangian and the
Hamiltonian of the system. In the following we use the former to get the relation between
$\boldsymbol{\pi}_j$ and $\dot{\boldsymbol{\alpha}}_j$ in order to construct
the Hamiltonian. Then Hamilton's equations along with Liouville's theorem
will be employed to get the time-evolution equation for the PDF of the system
so that its phases lie in a differential region of $\Gamma$ with centre placed at
$\mathfrak{s}(t)$.
First, the Lagrangian of the system can
be written as,\footnote{With $b$ and $B$ subscripts referring to the
\emph{bath} and \emph{Brownian} (colloidal) particles respectively.}
\begin{align}
 \mathcal{L}=&\mathcal{L}_b+\mathcal{L}_B,\label{eq:a-3}\\
 \mathcal{L}_b=&\sum_{i=1}^n\frac{1}{2}m_b\dot{\mathbf{r}}_j\cdot\dot{\mathbf{r}}_j-\mathcal{U}(\mathbf{r}^N,\mathbf{R}^n,\boldsymbol{\alpha}^n),\notag\\
 \mathcal{L}_B=&\sum_{i=1}^N\frac{1}{2}
 \left(m\dot{\mathbf{R}}_j\cdot\dot{\mathbf{R}}_j+\dot{\boldsymbol{\alpha}}_j\cdot(\boldsymbol{\Xi}_j^\top\mathbb{I}'_j\boldsymbol{\Xi}_j) \dot{\boldsymbol{\alpha}}_j\right)-V(\mathbf{R}^N,\boldsymbol{\alpha}^N)\notag,
\end{align}
with $V(\mathbf{R}^N,\boldsymbol{\alpha}^N)$ the potential energy due to short-range interactions exclusively between colloidal particles, and
\begin{equation}
 \mathcal{U}(\mathbf{r}^n,\mathbf{R}^N,\boldsymbol{\alpha}^N) = U(\mathbf{r}^n)+\sum_{\mu=1}^N\mathfrak{u}_\mu(\mathbf{r}^N,\mathbf{R}_\mu,\boldsymbol{\alpha}_\mu),\label{eq:a-4}
\end{equation}
the short-range intermolecular potential energy coming from the interaction
between bath particles, $U(\mathbf{r}^n)$, and the interaction of each
colloidal particle with the whole bath,
$\mathfrak{u}_\mu(\mathbf{r}^N,\mathbf{R}_\mu), \forall \mu=1,\dots,N$. From
these equations we obtain
\begin{align}
 \mathbf{p}_j=\frac{\partial \mathcal{L}}{\partial \dot{\mathbf{r}}_j}=m_b\dot{\mathbf{r}_j},\quad
 \mathbf{P}_j=\frac{\partial \mathcal{L}}{\partial \dot{\mathbf{R}}_j}=m\dot{\mathbf{R}_j},\quad
 \boldsymbol{\pi}_j=\frac{\partial \mathcal{L}}{\partial \dot{\boldsymbol{\alpha}}_j}= (\boldsymbol{\Xi}_j^\top\mathbb{I}'_j\boldsymbol{\Xi}_j)\dot{\boldsymbol{\alpha}_j},
 \label{eq:a-5}
\end{align}
Therefore, $\boldsymbol{\pi}_j$ can be easily related with the angular momentum $\mathbf{L}_j$ via\cite{condiff_brownian_1966,condiff_transport_1969}
\begin{equation}
\boldsymbol{\pi}_j = (\boldsymbol{\Xi}_j^\top\mathbb{I}'_j\boldsymbol{\Xi}_j)\dot{\boldsymbol{\alpha}}_j
\equiv
\boldsymbol{\Lambda}_j{\mathbf{L}}_j,
 \label{eq:a-6}
\end{equation}
which will be used to perform the appropriate change of variables later on.
We now write the Hamiltonian function of the system as
\begin{align}
 \mathcal{H} =& \mathcal{H}_b + \mathcal{H}_B \label{eq:a-7}\\
 \mathcal{H}_b =& \sum_{i=1}^n\frac{\mathbf{p}_j\cdot\mathbf{p}_j}{2m_b}+\mathcal{U}(\mathbf{r}^N,\mathbf{R}^n,\boldsymbol{\alpha}^n)\notag\\
 \mathcal{H}_B =&\sum_{i=1}^N\left(\frac{\mathbf{P}_j\cdot\mathbf{P}_j}{2m}+\frac{1}{2}\boldsymbol{\pi}_j\cdot(\boldsymbol{\Xi}_j^\top\mathbb{I}'_j\boldsymbol{\Xi}_j)^{-1}\boldsymbol{\pi}_j
 \right)+V(\mathbf{R}^N,\boldsymbol{\alpha}^n).\notag
\end{align}
According to Liouville's theorem,
the $n+N$ particle distribution function,
$\varrho^{(n+N)}(\mathbf{x}^n,\mathbf{X}^N;t)$, will evolve according to
\begin{equation}
 \partial_t \varrho^{(n+N)}(\mathbf{x}^n,\mathbf{X}^N,\boldsymbol{\Omega}^N;t)+\left(i\mathfrak{L}_b+i\mathfrak{L}_B^T+i\mathfrak{L}_B^R\right)\,\varrho^{(n+N)}(\mathbf{x}^n,\mathbf{X}^N;t)=0
 \label{eq:a-8}
\end{equation}
with the Liouvillian, $\mathfrak{L}\doteq \mathfrak{L}_b+\mathfrak{L}_B^T+\mathfrak{L}_B^R$,
\begin{align}
i\mathfrak{L}_b =& \sum_{j=1}^n\left(\frac{\mathbf{p}_j}{m_b}\cdot\frac{\partial }{\partial \mathbf{r}_j}+
\mathbf{f}_j\cdot\frac{\partial }{\partial \mathbf{p}_j}\right)\label{eq:a-9}\\
i\mathfrak{L}_B^T =& \sum_{j=1}^N\left(\frac{\mathbf{P}_j}{m}\cdot\frac{\partial }{\partial \mathbf{R}_j}+
\mathbf{F}_j\cdot\frac{\partial }{\partial \mathbf{P}_j}\right)\notag\\
i\mathfrak{L}_B^R =& \sum_{j=1}^N\left(\dot{\boldsymbol{\alpha}}_j\cdot\frac{\partial }{\partial \boldsymbol{\alpha}_j}+
\dot{\boldsymbol{\pi}}_j\cdot\frac{\partial }{\partial \boldsymbol{\pi}_j}\right)\notag
\end{align}
where
$\mathbf{f}_j=-\frac{\partial}{\partial\mathbf{r}_j}(U+\sum_{\mu=1}^N\mathfrak{u}_\mu)$
and $\mathbf{F}_j=-\frac{\partial}{\partial \mathbf{R}_j}(V+\mathfrak{u}_j)$
are the instantaneous forces acting on bath and colloidal particles,
respectively. From a conceptual point of view,\cite{dahler_kinetic_1963} the
quantities ${\boldsymbol{\alpha}}_j$ and ${\boldsymbol{\pi}}_j$ are less
convenient than the angular velocities and momenta, $\boldsymbol{\omega}_j$
and $\mathbf{L}_j$.
This can be easily checked by
considering the rotational kinetic energy
\begin{align}
K_{R}= \sum_{j=1}^N \frac{\dot{\boldsymbol{\alpha}}_j}{2}\cdot(\boldsymbol{\Xi}_j^\top\mathbb{I}'_j\boldsymbol{\Xi}_j) \dot{\boldsymbol{\alpha}}_j \equiv &\ \sum_{j=1}^N \frac{\boldsymbol{\omega}_j}{2}\cdot \mathbb{I}\boldsymbol{\omega}_j\label{a-10}\\
K_{R}= \sum_{j=1}^N \frac{\boldsymbol{\pi}_j}{2}\cdot(\boldsymbol{\Xi}_j^\top\mathbb{I}'_j\boldsymbol{\Xi}_j)^{-1}\boldsymbol{\pi}_j
\equiv&\ \sum_{i=1}^N \frac{\mathbf{L}_j}{2}\cdot(\mathbb{I}^{-1}\mathbf{L}_j).\notag
\end{align}
Transforming now Equation (\ref{eq:a-8}) into an equivalent equation for
\begin{align}
 F^{(n+N)}(\mathbf{x}^n,\mathbf{X}^N,\boldsymbol{\alpha}^N,\mathbf{L}^N;t)=&\left|
 \frac{\partial(\mathbf{x}^n,\mathbf{X}^N,\boldsymbol{\alpha}^N,\boldsymbol{\pi}^N)}{\partial (\mathbf{x}^n,\mathbf{X}^N,\boldsymbol{\alpha}^N,\mathbf{L}^N)}
 \right|\varrho^{(n+N)}(\mathbf{x}^n,\mathbf{X}^N,\boldsymbol{\Omega}^N;t)\label{eq:a-11}\\
 =&\prod_{j}\sin\theta_j\,\varrho^{(n+N)}(\mathbf{x}^n,\mathbf{X}^N,\boldsymbol{\Omega}^N;t)\notag
\end{align}
gives
\begin{align}
 \partial_t F^{(n+N)}(\mathbf{x}^n,\mathbf{X}^N,\boldsymbol{\Omega}^N;t)+\left(i\mathfrak{L}_b+i\mathfrak{L}_B^T+i\widetilde{\mathfrak{L}}_B^R\right)\,F^{(n+N)}(\mathbf{x}^n,\mathbf{X}^N,\boldsymbol{\Omega}^N;t)=0
 \label{eq:a-12}
\end{align}
with\cite{condiff_brownian_1966,condiff_transport_1969,dahler_kinetic_1963}
\begin{equation}
i\widetilde{\mathfrak{L}}_B^R = \sum_{j=1}^N\dot{\boldsymbol{\alpha}}_j \cdot \hat{\partial}_{\boldsymbol{\alpha}_j}+\mathbf{T}_j\cdot\frac{\partial }{\partial \mathbf{L}_j}
\label{eq:a-13}
\end{equation}
where the operator\cite{condiff_brownian_1966}
$\hat{\partial}_{\boldsymbol{\alpha_j}}\doteq
(\csc\theta_j\frac{\partial}{\partial \theta_j}\sin\theta_j,
\frac{\partial}{\partial \phi_j}, \frac{\partial}{\partial\chi_j})^\top$,
deduced by using the chain rule\cite{risken_fokker-planck_1996},
has been introduced. In the latter equation, $\mathbf{T}_j = \mathbf{N}_j -
\boldsymbol{\omega}_j\times\mathbf{L}_j$ denotes the net torque acting on the
$j$-th colloidal particle, with\cite{condiff_brownian_1966}
\begin{align}
 \mathbf{N}_j &= - \boldsymbol{\Lambda}_j^{-1}\frac{\partial}{\partial \boldsymbol{\alpha}_j}\left[V(\mathbf{R}^N,\boldsymbol{\alpha}^n)+\mathfrak{u}_j(\mathbf{r}^N,\mathbf{R}_j,\boldsymbol{\alpha}_j)\right]
 \label{eq:a-14}
\end{align}
the torque due to intermolecular interactions along the principal axes of
inertia. At this point, it is necessary to introduce the angular-gradient
operator (also known as orientational
gradient\cite{peters_smoluchowski_2000})
\begin{align}
 \frac{\partial }{\partial \boldsymbol{\Phi}_j}&\doteq
 \boldsymbol{\Lambda}_j^{-1}\frac{\partial }{\partial \boldsymbol{\alpha}_j}\equiv \hat{\partial}_{\boldsymbol{\alpha}_j}\cdot(\boldsymbol{\Lambda}^\top)^{-1}\label{eq:a-15}
\end{align}
such that $\frac{\partial }{\partial \bPhi}= \mathbf{e}_x\frac{\partial
}{\partial \Phi_{x}}+\mathbf{e}_y\frac{\partial }{\partial
\Phi_{y}}+\mathbf{e}_z\frac{\partial }{\partial \Phi_{z}}$, where
$\mathbf{e}_i$ is the unitary vector along axis $i\in\{x,y,z\}$ of the
Cartesian frame $\mathfrak{B}$, and
\begin{align}
 \frac{\partial }{\partial \Phi_{x}^j}&= \cos\chi_j\frac{\partial}{\partial \theta_j}
 +\csc\theta_j\cos\chi_j\frac{\partial }{\partial \phi_j}
 -\cot\theta_j\sin\chi_j\frac{\partial }{\partial \chi_j}\label{eq:a-16}\\
 \frac{\partial }{\partial \Phi_y^j}&=
 -\sin\chi_j\frac{\partial}{\partial \theta_j}
 +\csc\theta_j\sin\chi_j\frac{\partial }{\partial \phi_j}
 -\cot\theta_j\cos\chi_j\frac{\partial }{\partial \chi_j}
 \notag\\
 \frac{\partial }{\partial \Phi_z^j}&=\frac{\partial }{\partial\chi_j}.\notag
\end{align}
It is worth mentioning here that the derivative operators $(\frac{\partial
}{\partial \Phi_{x}},\frac{\partial }{\partial \Phi_{y}},\frac{\partial
}{\partial \Phi_{z}})$ are the generators of rotations of a rigid body about
the body-fixed Cartesian frame\cite{gray_theory_1984}. This results in
\begin{align}
i\widetilde{\mathfrak{L}}_B^R=\sum_{j=1}^N \frac{\partial }{\partial \boldsymbol{\Phi}_j}\cdot\boldsymbol{\omega}_j +  \left(\mathbf{N}_j+\mathbf{L}_j\times\boldsymbol{\omega}_j\right)\cdot\frac{\partial }{\partial\mathbf{L}_j}
\label{eq:a-17}.
\end{align}
Now we can define $\bfm\doteq \text{diag}(m\mathbf{1},\mathbb{I})$
along with the vectors $\boldsymbol{\mathfrak{r}}_j\doteq
(\mathbf{R}_j,\boldsymbol{\Phi}_j)$,
$\boldsymbol{\mathfrak{p}}_j\doteq(\mathbf{P}_j,\mathbf{L}_j)$ and
$\boldsymbol{\mathfrak{f}}_j\doteq(\mathbf{F}_j,\mathbf{T}_j)$, and the
operators
$\boldsymbol{\nabla}_{\mathfrak{r}_j}\doteq(\frac{\partial}{\partial
\mathbf{R}_j},\frac{\partial }{\partial \boldsymbol{\Phi}_j})^\top$ and
$\boldsymbol{\nabla}_{\boldsymbol{\mathfrak{p}}_j}\doteq (\frac{\partial
}{\partial \mathbf{P}_j},\frac{\partial }{\partial \mathbf{L}_j})^\top$,
enabling us to rewrite Equation (\ref{eq:a-12}) in a more compact and convenient
way,
\begin{align}
 \partial_t F^{(n+N)}(t)+
 \sum_{j=1}^n\left(\frac{\mathbf{p}_j}{m_b}\cdot\frac{\partial }{\partial \mathbf{r}_j}+
\mathbf{f}_j\cdot\frac{\partial }{\partial \mathbf{p}_j}\right)\FnN(t)
 + \sum_{j=1}^N\left(\bnabla_{\bfr_j}\cdot \bfm^{-1}\bfp_j + \bff_j\cdot\bnabla_{\bfp_j}
 \right)\FnN(t)=0\label{eq:a-18},
\end{align}
where explicit dependence on the phase-space coordinates was omitted, but
recalled through the superscript $(n+N)$.
Following \citet{murphy_brownian_1972}, we introduce the
scaling quantity, $\blambda \doteq \bfm^{-1/2}$, so that
$\widetilde{\bfp}_j=\blambda\bfp_j$ and hence, the last term of the previous
equation becomes
\begin{align}
 \sum_{j=1}^N\blambda\left(\bnabla_{\bfr_j}\cdot \wbfp_j + \bff_j\cdot\bnabla_{\wbfp_j}
 \right)\FnN(t). \label{eq:a-19}
\end{align}
Substitution of Equation (\ref{eq:a-19}) into Equation (\ref{eq:a-18}) the results in
an equation resembling Liouville's equation for spherical colloidal
particles. Such a result provides a description of the time evolution of the
full system. However, our interest rests exclusively on colloidal particles.
Thus, our aim is to get the time-evolution equation for the $N$-particle
distribution,
\begin{align}
 \fN(t) \doteq \int d\bx^n  \FnN(t).\label{eq:a-20}
\end{align}
For this purpose, Zwanzig's projection technique can be applied as in the
case of spherical colloidal particles\cite{deutch_molecular_1971,
murphy_brownian_1972, wilemski_derivation_1976, ermak_brownian_1978}.
Following the work of \citet{murphy_brownian_1972}, for the arbitrary initial
state at $t=-t_I$ we choose one where the bath particles are in equilibrium with
the instantaneous positions of the colloidal particles.
This means that
\begin{equation}
  \FnN(\bx^n,\bX^N,\bOmega^N;-t_I)=\rho_n^\dag(\bx^n)\,\fN(\bX^N,\bOmega^N;-t_I)\label{eq:a-21},
\end{equation}
with $\rho_n^\dag$ the canonical distribution of the $n$ bath particles
in the instantaneous potential created by the colloidal particles. The last step before integrating Equation (\ref{eq:a-18})
to remove the dependence on fast variables involves the definition of the
projection operator,
\begin{equation}
 \hat{\mathcal{P}}\doteq \rho_n^\dag(\bx^n)\int d\bx^n\,,
\end{equation}
and its complementary operator $\hat{\mathcal{Q}}=1-\hat{\mathcal{P}}$. Thus,
the integration of Equation (\ref{eq:a-18}) over the fast variables $\bx^n$ is
equivalent to applying $(\rho_n^\dag)^{-1}\hat{\mathcal{P}}$ on both sides of
such an equation. Although considerable algebraic manipulations are
required,
we can follow the work of \citet{murphy_brownian_1972} and
\citet{lebowitz_microscopic_1965} step by step to finally reach the desired
time-evolution equation for the projected distribution function $\fN$,
\begin{align}
 \partial_t\fN(t)+\sum_{j=1}^N&\blambda\left(\bnabla_{\bfr_j}\cdot \wbfp_j+\bfF_j\cdot\bnabla_{\wbfp_j}\right)\fN(t) \label{eq:a-23}\\
 =& \sum_{j,k=1}^N \blambda^2\,\bnabla_{\bfp_j}\cdot\int_{-t_I}^{t}dt'\bgamma_{jk}(\blambda;t,t')\left(\bnabla_{\wbfp_k}+\beta\,\wbfp_k\right)\fN(t')\notag
\end{align}
where $\beta=1/k_BT$, where $k_B$ is the Boltzmann constant, $T$ is the
temperature imposed by the bath, and $\bfF$ is the equilibrium average force
and torque, i.e.
\begin{align}
  \bfF_j &\equiv
   \left\langle
  \begin{array}{l}
  \bF_j\\
  \bT_j
 \end{array}
  \right\rangle^\dag
  =\int d\bx^n \rho_n^\dag(\bx^n)\,\bff_j(\br^n,\bR^N,\balpha^N)
 \label{eq:a-24}\\
 &=-\left[
 \begin{array}{l}
  -\frac{\partial }{\partial \bR_j} V(\bR^N,\balpha^N)\\
  -\frac{\partial }{\partial \bPhi_j}V(\bR^N,\balpha^N)- \bomega_j\times\bL_j
 \end{array}
 \right]+\left\langle\bnabla_{\bfr_j}\mathfrak{u}(\br^n,\bR_j,\balpha_j)\right\rangle^\dag =
 \left[
 \begin{array}{l}
  -\frac{\partial }{\partial \bR_j} (V+\psi)\\
  -\frac{\partial }{\partial \bPhi_j}(V+\psi)- \bomega_j\times\bL_j
 \end{array} \right],
 \notag
\end{align}
with $\langle.\rangle^\dag$ the equilibrium average over the fast variables.
Equation (\ref{eq:a-24})
also includes the definition of the potential of mean force $\psi\doteq
\langle\mathfrak{u}\rangle^\dag$, i.e. the potential which gives rise to the
average (over all configurations of the $n$ bath molecules) force and torque
acting on the $j$th colloidal particle at any given configuration keeping all
the colloidal particles frozen.
Thus, the fluid-equilibrium average
force and torque includes the contribution of a postulated vacuum
colloid-colloid interaction potential, $V$, and the solvent contribution to
the total force and torque, $\psi$. This combination in turn results in a
solvent-averaged potential of mean force, $\widetilde{V}\doteq V+\psi$, which
can be obtained from a given physical model, e.g. the DLVO theory for the
interaction of charged colloidal particles\cite{snook_langevin_2006} or the
ten Wolde-Frenkel potential\cite{wolde_enhancement_1997}.
Finally, the tensor $\bgamma_{jk}$ is given by
\begin{align}
 \bgamma_{jk}(\blambda;t,t')&=\left\langle \bff_j(t)\otimes e^{i(t'-t)\hat{\mathcal{Q}}\mathfrak{L}}(\bff_k(t)-\bfF_k)\right\rangle^\dag
 \label{eq:a-25}\\
 &\equiv\left\langle
 \begin{array}{ccc}
  \bF_j(t)\otimes e^{i(t'-t)\hat{\mathcal{Q}}\mathfrak{L}}(\bF_k(t)-\boldsymbol{\mathcal{F}}_k)
&& \bF_j(t)\otimes e^{i(t'-t)\hat{\mathcal{Q}}\mathfrak{L}}(\bT_k(t)-\boldsymbol{\mathcal{T}}_k)\\
  \bT_j(t)\otimes e^{i(t'-t)\hat{\mathcal{Q}}\mathfrak{L}}(\bF_k(t)-\boldsymbol{\mathcal{F}}_k)
&& \bT_j(t)\otimes e^{i(t'-t)\hat{\mathcal{Q}}\mathfrak{L}}(\bT_k(t)-\boldsymbol{\mathcal{T}}_k)
  \end{array}
 \right\rangle^\dag\notag.
\end{align}
where $\boldsymbol{\mathcal{F}}\doteq\langle\bF_k\rangle^\dag$ and
$\boldsymbol{\mathcal{T}}\doteq\langle\bT\rangle^\dag$. The behaviour of Equation
(\ref{eq:a-23}) as the product $m_b\,\bfm^{-1}$ vanishes can be obtained by
simply letting $\blambda\rightarrow \mathbf{0}$. In such a limit, the friction
tensor $\bgamma$ can be approximated by the first term of a multi-power
series in $\blambda$,\cite{murphy_brownian_1972}
\begin{align}
 \bgamma_{jk}(\blambda;t,t')\sim \left\langle
 \bff_j(t)\otimes (\bff_k(t')-\bfF_k)
 \right\rangle\label{eq:a-26}
\end{align}
which yields
\begin{align}
 \partial_t\fN(t)+\blambda\sum_{j=1}^N&\left(\bnabla_{\bfr_j}\cdot \wbfp_j+\bfF_k\cdot\bnabla_{\wbfp_j}\right)\fN(t) \label{eq:a-27}\\
 =& \sum_{j,k=1}^N \blambda^2\,\bnabla_{\bfp_j}\cdot\int_{-t_I}^{t}dt'\left\langle
 \bff_j(t)\otimes (\bff_k(t')-\bfF_k)
 \right\rangle\left(\bnabla_{\wbfp_k}+\beta\,\wbfp_k\right)\fN(t'). \notag
\end{align}
The last step, the ``Markovianization'' of this time-evolution equation with
memory, is the most controversial
one\cite{snook_langevin_2006,mazo_theory_1969,mazur_molecular_1970,bocquet_microscopic_1997}
as it requires the assumption that $\fN(t')$ is very slowly varying compared
to the correlation $\langle \bff_j(t)\otimes\bff_k(t) \rangle^\dag$, so
that\cite{murphy_brownian_1972}
\begin{equation}
 \int_{-t_I}^{t}dt'\left\langle
 \bff_j(t)\otimes (\bff_k(t')-\bfF_k)
 \right\rangle\left(\bnabla_{\wbfp_k}+\beta\,\wbfp_k\right)\fN(t')\sim \int_{-\infty}^{t}dt'\left\langle
 \bff_j(t)\otimes (\bff_k(t')-\bfF_k)
 \right\rangle\left(\bnabla_{\wbfp_k}+\beta\,\wbfp_k\right)\fN(t). \label{eq:a-28}
\end{equation}
Such an approximation is widely known to produce an incorrect description of
the velocity correlation function if the colloidal particles have a similar
density to that of the bath particles\cite{snook_langevin_2006,
mazo_theory_1969, bocquet_microscopic_1997, mazur_molecular_1970}. Equation
(\ref{eq:a-28})
gives an exponential decay, $\sim e^{-t}$, for the velocity correlation at
large times, while with memory, such a decay is
algebraic\cite{snook_langevin_2006}, $\sim t^{-3/2}$.
Nevertheless, it has also been pointed out that such long-time tails
are very small compared to the exponential component predicted under the
Markovianized theory\cite{snook_langevin_2006}. Although these could be
reasons to avoid this critical step, the advantages of getting an FPE are
significant. In contrast, the approximation is completely valid when both
$m_b/m$ and $N_{\text{Kn}}\doteq r_0/R_0$ (the Knudsen number, with $r_0$
being a characteristic length scale for fluid intermolecular interactions and
$R_0$ is a characteristic colloidal particle length scale) are considered
very small\cite{peters_fokker-planck_1999}. Nevertheless, it was argued by
\citet{bocquet_microscopic_1997} that under such circumstances sedimentation
of colloidal particles could occur. This seems to restrict the applicability
of the resultant theory to microgravity scenarios. However, it is not
clear at all what the actual significance of these algebraic long-time tails is
and each case should be judged on its own merits\cite{mazo_theory_1969}. Thus Equation
(\ref{eq:a-28}) comprises, on the one hand, an uncontrolled approximation. On
the other hand, it has been extensively used in statistical
mechanics\cite{curtiss_kinetic_1957, wolynes_dynamical_1977,
peters_fokker-planck_1999, hernandez-contreras_brownian_1996}. For instance,
the same hypothesis underlies a recent derivation of a unified DDFT to
include inertia and HIs\cite{goddard_unification_2013}, and is also involved
in modern theories describing nucleation of colloidal systems and
macromolecules\cite{lutsko_dynamical_2012,lutsko_classical_2013,duran-olivencia_mesoscopic_2015,lutsko_two-parameter_2015}.
Moreoever, the results obtained are in perfect agreement with experiments and
simulations, corroborating the smallness of the error related to Equation
(\ref{eq:a-28}) when it comes to describing systems of interacting colloidal
particles. More akin to the problem at hand, the hypothesis is tacitly
assumed within the seminal work of \citet{dickinson_brownian_1985}, where a
generalised algorithm to simulate protein diffusional problems is proposed.
Therefore, while we cannot really justify such an assumption in a rigorous
manner it does, nevertheless, represents the state-of-the-art in modelling
colloidal systems. With this in mind, we can finally obtain the FPE related
to Equation (\ref{eq:a-27}) when Equation (\ref{eq:a-28}) is taken into consideration,
\begin{align}
 \partial_t\fN(t)+\sum_{j=1}^N\left(\bnabla_{\bfr_j}\cdot \bfm^{-1}\bfp_j+\bfF_j\cdot\bnabla_{\bfp_j}\right)\fN(t) = \sum_{j,k=1}^N \,\bnabla_{\bfp_j}\cdot\bGamma_{jk}(\bfr^N)\left(\bfp_k+k_BT\bfm\,\bnabla_{\bfp_k}\right)\fN(t),
 \label{eq:a-29}
\end{align}
with the translational, rotational and coupled translational-rotational friction tensors,
\begin{equation}
\bGamma_{jk}(\bfr^N)
\equiv
 \left(
 \begin{array}{cc}
  \bGamma_{jk}^{TT} & \bGamma_{jk}^{TR}\\
  \bGamma_{jk}^{RT} & \bGamma_{jk}^{RR}\\
 \end{array}
 \right)
\doteq \beta \bfm^{-1}\int_{0}^{\infty}ds\,\left\langle
\bff_j(t)\otimes (\bff_k(t-s)-\bfF_k)
 \right\rangle. \label{eq:a-30}
\end{equation}

\subsection{Equations of motion}
 The FPE previously derived can be rewritten in the less compact but more explicit form,
\begin{align}
\partial_t \fN(t)+&\sum_{j=1}^N
 \bnabla_{\bfr_j}\cdot
 \left(\bfm^{-1}\bfp_j\fN(t)\right)
 -
  \bnabla_{\bfp_j}
 \cdot \left[\left(\bfF_j-\sum_{k=1}^N\bGamma_{jk}(\bfr^N)\bfp_k
 \right)\fN(t)\right]\label{eq:a-31}\\
 &= \sum_{j,k=1}^N
\left[
\bnabla_{\bfp_j} \otimes
\bnabla_{\bfp_k}
\right]:\left(k_BT\bfm\,\bGamma_{jk}(\bfr^N)\fN(t)\right)\notag
\end{align}
which is equivalent to the system of
SDEs\cite{risken_fokker-planck_1996,kampen_stochastic_2011,berendsen_simulating_2007}
\begin{align}
 \dot{\bfr}_j(t)=&\ \bfm^{-1}\bfp_j(t)\label{eq:a-32}\\
 \dot{\bfp}_j(t)=&\ \bfF_j-\sum_{k=1}^N\bGamma_{jk}(\bfr^N)\bfp_k + \sum_{k=1}^N\mathbf{A}_{jk}\boldsymbol{\xi}_k(t)\notag
\end{align}
where $\bxi_j=(\mathbf{f}_j,\mathbf{t}_j)^\top$ is a 6-dimensional Gaussian
white noise representing the random forces, $\mathbf{f}_j$, and torques,
$\mathbf{t}_j$, acting upon the $j$-th particle, such that $\langle
\xi_j^a(t)\rangle=0$ and $\langle \xi_j^a(t)\xi_k^b(t')\rangle  =
2\delta_{jk}\delta^{ab}\delta(t-t')$, where $\langle.\rangle$ refers to the
average over an ensemble of the white-noise realisations. The strength of
these random forces and torques is given by the tensor $\mathbf{A}_{jk}$
which obeys the fluctuation-dissipation relation,
\begin{equation}
 k_BT \bfm\,\bGamma_{jk}(\bfr^N) = \sum_{l=1}^N\mathbf{A}_{jl}(\bfr^N)\mathbf{A}_{kl}(\bfr^N)
\label{eq:a-33},
\end{equation}
and
\begin{align}
 \mathbf{A}_{jk}\bxi_k\equiv
 \begin{pmatrix}
\mathbf{A}_{jk}^{TT}&\mathbf{A}_{jk}^{TR}\\
\mathbf{A}_{jk}^{RT}&\mathbf{A}_{jk}^{RR}
 \end{pmatrix}
 \begin{pmatrix}
  \mathbf{f}_{k}\\
  \mathbf{t}_{k}
 \end{pmatrix}.
 \label{eq:a-34}
\end{align}
Coming back to the expanded notation, the system of equations (\ref{eq:a-32}) becomes
\begin{align}
 \frac{d\bR_j}{dt}=&\,\frac{1}{m}\bP_j,\label{eq:a-35}\\
 \frac{d\bPhi_j}{dt}=&\,\mathbb{I}^{-1}\bL_j\label{eq:a-36},\\
 \frac{d\bP_j}{dt}=&\, -\frac{\partial}{\partial \bR_j}\widetilde{V}(\bR^N,\balpha^N)-\sum_{k=1}^N\left(\bGamma_{jk}^{TT}\bP_k+\bGamma_{jk}^{TR}\bL_k\right)+\sum_{k=1}^N\mathbf{A}_{jk}^{TT}\mathbf{f}_k(t)+\mathbf{A}_{jk}^{TR}\mathbf{t}_k(t)
 \label{eq:a-37}\\
 \frac{d\bL_j}{dt}\equiv\mathbb{I}\frac{d\bomega_j}{dt}=&
 -\frac{\partial}{\partial \bPhi_j}\widetilde{V}(\bR^N,\balpha^N)-\bomega_j\times\bL_j
 \,
 -\sum_{k=1}^N\left(\bGamma_{jk}^{RT}\bP_k+\bGamma_{jk}^{RR}\bL_k\right)+\sum_{k=1}^N\mathbf{A}_{jk}^{RT}\mathbf{f}_k(t)+\mathbf{A}_{jk}^{RR}\mathbf{t}_k(t)
 \label{eq:a-38}
\end{align}
with Equation (\ref{eq:a-36}) equivalent to the relation
$\bomega_j=\bLambda_j^\top\dot{\balpha_j}$, as can be verified by using
Equation (\ref{eq:a-15}). Equations~(\ref{eq:a-35}-\ref{eq:a-38}), a much
less convenient representation of the rotational-translational Langevin
equation (\ref{eq:a-32}), have been introduced in the studies by
\citet{wolynes_dynamical_1977},
\citeauthor{dickinson_brownian_1985}\cite{dickinson_brownian_1985,
dickinson_brownian_1985-1} and \citet{hernandez-contreras_brownian_1996} but
these authors started from postulated equations instead of the detailed
microscopic derivation from the full system of bath and colloidal particles
offered here.

\section{On the rapid relaxation of the fluid: neglecting inertia in the bath}\label{app:neglecting-bath-inertia}

Here we address the question of when the inertia of the fluid bath can be
neglected while having finite viscous forces.
With this aim we will make use of some of the results derived by
\citet{peters_fokker-planck_1999} in his study on the FPE for
coupled rotational and translational motions of structured Brownian particles.
The main conclusion, the rapid relaxation of the
fluid bath, is ultimately connected with the assumption of negligible inertial
effects in the fluid bath whilst considering inertial effects of the
colloids.

In a very detailed study, \citet{peters_fokker-planck_1999} applied the
multiple time-scale expansion to the derivation of the FPE for
arbitrary-shape colloids. With this method \citeauthor{peters_fokker-planck_1999}
showed that Equation (\ref{eq:a-29}) is the formal time-evolution equation for the
distribution function, up to $(m/M)^3$, when both $m/M$ and $N_{\text{Kn}}$
are considered small. It was also argued that the rapid relaxation of the
fluid depends upon $N_{\text{Kn}}$ for a system with $m/M$ small. If we now
make use of the fact that $m \sim r_0^3\rho_b$ and $M\sim R_0^3\rho_B$, the
conditions for the FPE would to be a good description of the colloidal system
can be reduced to requiring that $\rho_b/\rho_B$ must be small, which is
indeed the condition pointed out by \citet{bocquet_microscopic_1997} and many
others\cite{roux_brownian_1992,hauge_fluctuating_1973,hinch_application_1975,
masters_time-scale_1986, michaels_long-time_1975}. In the following we
analyse whether or not this is possible while the ratio between inertia in
the fluid bath and viscous forces is small. That is, whether is possible to
have a low Reynolds number for the fluid bath
along with the condition on the
ratio between bath and colloidal densities.
In such a case, it would be justified to neglect
inertial effects in the bath.

To this end, we consider the case of solid and spherical colloids (far from
walls) so that the coupling components of the friction and diffusion tensors
vanish, and neglect HIs for the moment. If the radius of the colloidal
particles is denoted by $R_0$, the friction tensor takes the simple
form\cite{dickinson_brownian_1985},
\begin{align}
 \bGamma_{jk}^{TT} = \gamma_{T}\delta_{jk}\,\mathbf{1},\quad \bGamma_{jk}^{RR}=\gamma_{R}\delta_{jk}\,\mathbf{1},\quad \bGamma_{jk}^{TR}=\bGamma_{jk}^{RT}=0,\label{eq:b-1}
\end{align}
with  $\gamma_{T}=6\pi\eta\, R_0/M$ and
$\gamma_{R}=8\pi\eta\,R_0^3/I=20\pi\eta\,R_0/M$, and $\eta$ the dynamic
viscosity, satisfying the Stokes-Einstein
formula\cite{dickinson_brownian_1985}. For the sake of generality, the
friction components will be considered equally important, i.e.
$\gamma_{T}/\gamma_{R}\sim\mathcal{O}(1)$. For this reason, we can define the
following two time scales
\begin{align}
 t_0^T = \left(\frac{MR_0^2\,\gamma_{TT}}{k_BT}\right),\quad t_0^R=\left(\frac{MR_0^2\,\gamma_{RR}}{k_BT}\right)
 \label{eq:b-2},
\end{align}
which are indeed of the same order-of-magnitude.

 Considering the natural physical scales of the system, the following
dimensionless translational and rotational variables (denoted by an asterisk)
are pertinent
\begin{align}
\begin{array}{ccccc}
 t = t_0^T\,t*,\ &\ \bR_j = R_0\,\bR^*_j,\ &\ \bP_j =  \frac{k_BT}{\gamma_TR_0}\,\bP_j^*, \ &\ \mathbf{F}_j = \frac{k_BT}{R_0}\,\mathbf{F}_j^*,\ &\ \mathbf{F}_{j,\text{noise}}= \frac{k_BT}{R_0}\,\mathbf{F}_{j,\text{noise}}^*,\ \\
 t = t_0^R\,t*,\ &\ \bomega_j = \frac{k_BT}{MR_0^2\gamma_R}\bomega^*_j,\ &\ \bL_j =  \frac{k_BT}{\gamma_R}\,\bL_j^*, \ &\ \mathbf{T}_j = k_BT\,\mathbf{T}_j^*,\ &\ \mathbf{T}_{j,\text{noise}}= k_BT\,\mathbf{T}_{j,\text{noise}}^*,\ \\
\end{array}\label{eq:b-3}
\end{align}
which, when applied in Equations (\ref{eq:a-35})-(\ref{eq:a-38}), yield the
dimensionless equations of motion,
\begin{align}
 \frac{d\bR_j^*}{dt^*}=&\,\bP_j^*,\label{eq:b-4}\quad  \frac{d\bPhi_j^*}{dt^*}=\,\bomega_j^*\\
 \frac{d\bP_j^*}{dt^*}=&\,\gamma_T^{*^2}\left(\mathbf{F}_j^*-\bP_j^*+\mathbf{F}_{j,\text{noise}}^*(t)\right)\notag\\
 \frac{d\bL_j^*}{dt^*}=&\,\gamma_R^{*^2}\left(\mathbf{T}_j^*-\bL_j^*+\mathbf{T}_{j,\text{noise}}^*(t)\right)\notag
\end{align}
along with the definitions
\begin{align}
 \gamma_T^{*}=\gamma_T R_0\sqrt{\frac{M}{k_BT}},\quad \gamma_R^*=\gamma_{R} R_0\sqrt{\frac{M}{k_BT}}.
 \label{eq:b-5}
\end{align}
What we wish to test is whether or not it is possible to have finite viscous
forces, i.e. $\gamma_T^{*}\sim\mathcal{O}(1)$, such that at the same time
inertia forces in the fluid bath are negligible.
For this purpose both translational and rotational
Reynolds numbers must be small. They can be defined as\cite{happel_low_1981},
\begin{align}
\text{Re}^T=\frac{U\,R_0\,\rho_b}{\eta},\quad \text{Re}^R=\frac{\Omega\,R_0^2\,\rho_b}{\eta},
\label{eq:b-6}
\end{align}
respectively. The quantities $U$ and $\Omega$ represent typical linear and angular velocity scales, which can be obtained from the nondimensionalised momenta,
\begin{equation}
U=\frac{k_BT}{MR_0\gamma_{T}},\quad \Omega=\frac{k_BT}{MR_0^2\gamma_{R}}.\label{eq:b-7}
\end{equation}
Making use of (\ref{eq:b-2}), (\ref{eq:b-3}), (\ref{eq:b-5}) and (\ref{eq:b-7}) into (\ref{eq:b-6}),  we finally reach
\begin{equation}
 \text{Re}^T=\frac{9}{2\gamma_{T}^{*^2}}\frac{\rho_b}{\rho_B},\quad \text{Re}^R=\frac{15}{\gamma_{R}^{*^2}}\frac{\rho_b}{\rho_B}\label{eq:b-8}.
\end{equation}
As already noted, the regime we consider involves,
$\gamma_T^{*}\sim\gamma_{R}^*\sim\mathcal{O}(1)$ and
$\text{Re}^T\sim\text{Re}^R\ll1$, which is undoubtedly satisfied when
$\rho_b/\rho_B\ll 1$. Then, neglecting inertia forces in the fluid bath
is consistent with the separation of time scales we already assumed to obtain the FPE.
Further details of the physical interpretation and consequences of this limiting condition were given in Section
\ref{app:projection-operator-tech}.


\section*{References}\vspace{-0.5cm}
\bibliographystyle{spmpscinat}

\end{document}